\pdfoutput=1
\documentclass[12pt]{article}
\usepackage[
top = 2.5cm, 
bottom = 2.5cm,  
left = 2.55cm,  
right = 2.55cm]{geometry}

\usepackage{subfigure}
\usepackage{caption}
\usepackage{latexsym} 
\usepackage{verbatim}  
\usepackage{tikz} 
\usetikzlibrary{matrix}
\usepackage{color}
\usepackage{graphicx,amssymb,amsfonts,amsmath,amssymb,amscd,amstext, mathrsfs}
\usepackage{graphicx} 
\usepackage{dsfont}
\usepackage{xcolor}
\usepackage{amsmath}
\usepackage{empheq}
\usepackage{cite}

\numberwithin{equation}{section}

\newlength\dlf

\newcommand{\bw}{\begin{widetext}}
\newcommand{\ew}{\end{widetext}}
\newcommand{\bea}{\begin{eqnarray}}
\newcommand{\eea}{\end{eqnarray}}
\newcommand{\be}{\begin{equation}}
\newcommand{\ee}{\end{equation}}

\renewcommand{\bar}[1]{\overline{#1}}

\newcommand{\<}{\langle}
\renewcommand{\>}{\rangle}

\renewcommand{\cal}{\mathcal}
\newcommand{\CO}{\mathcal{O}}

\DeclareFontShape{OT1}{cmr}{mx}{n}{<->cmr10}{}
\newcommand{\titlefont}{\fontseries{mx}\selectfont}

\def\frac#1#2{{#1\over #2}}

\usepackage{jheppub}

\begin{document}

\begin{titlepage}

\begin{flushright} 
\end{flushright}

\begin{center} 

\vspace{0.35cm}

{\fontsize{21.5pt}{0pt}{\titlefont 
Giving Hamiltonian Truncation a Boost
}}

\vspace{1.6cm}  

{{Hongbin Chen$^1$, A. Liam Fitzpatrick$^1$,  Emanuel Katz$^1$,  Yuan Xin$^{2}$}}

\vspace{1cm} 

{{\it
$^1$Department of Physics, Boston University, 
Boston, MA  02215, USA
\\
\vspace{0.1cm}
$^2$Department of Physics, Yale University, New Haven, CT 06520, USA
}}\\
\end{center}
\vspace{1.5cm}

{\noindent 
We study Hamiltonian truncation in boosted frames.  We consider the thermal and magnetic field deformations of the 2d Ising model using TCSA at finite momentum.  We find that even with moderate momenta, the spectrum and time-dependent correlation functions become significantly less dependent on the volume of the system.  This allows for a more reliable determination of infinite volume observables.   
}

\end{titlepage}

\tableofcontents


\section{Introduction}
Hamiltonian truncation \cite{Yurov:1989yu, Yurov:1991my, Anand:2020gnn} is a promising numerical technique for computing the non-perturbative data of a QFT\footnote{For some recent development and applications of Hamiltonian truncation, see \cite{Kikuchi:2022jbl, Emonts:2022vim, Horvath:2022zwx,  EliasMiro:2021aof, Lencses:2021igo, Cohen:2021erm, Cubero:2021sje, Bajnok:2021twm, Hogervorst:2021spa, Konechny:2020jym, Bajnok:2020kyk, Hodsagi:2020dqq, Elias-Miro:2020qwz,Dempsey:2021xpf, Konechny:2019rwb, Hogervorst:2018otc, Rutter:2018aog,  Elias-Miro:2017xxf, Rakovszky:2016ugs,Rychkov:2015vap, Rychkov:2014eea, Hogervorst:2014rta} and the references of \cite{James:2017cpc}.}.  
It provides both the energy spectrum as well as the Hamiltonian eigenstates, making it a natural approach for calculating the evolution of states in time, and thus of various time-dependent observables.  It also expresses all quantities solely in terms of the degrees of freedom of the QFT itself. Due to the need to discretize the QFT data, traditionally, Hamiltonian truncation is formulated in finite volume.  This is also sensible given Haag's theorem, which states that eigenstates of an interacting theory are orthogonal to the eigenstates of a free theory at infinite volume.  However, a relic of Haag's theorem persists even at finite volume because, in practice, one finds that the states can have significant sensitivity to the volume.  If observables have too much sensitivity to the volume, then it can become difficult in practice to robustly extract their infinite volume limit.
 
On the other hand,  it has long been known that lightcone quantization can be formulated directly in the infinite volume limit, and so while it might be too strong to say that it solves the problems of volume sensitivity, it at least converts them into a rather different form. 
 In this case, discretization is achieved by introducing some basis of discrete functions to parameterize the wavefunction. The basis is labeled by a dimensionless discrete parameter.  Crucially, in lightcone quantization, the Fock vacuum state does not mix with any other states, even in the presence of interactions. It thus remains trivial, and old-fashioned perturbation theory on the lightcone closely resembles calculations using Feynman diagrams.  Indeed, bubble diagrams are excluded and therefore there is no sensitivity to the volume at all.  However, there is a price for keeping the vacuum trivial:  the naive lightcone Hamiltonian needs to be corrected by integrating out lightcone zero modes.  In some cases, this correction is simple to calculate and is perturbative in the couplings, but in some other cases, the correction can be non-perturbative \cite{Fitzpatrick:2018ttk}.  In particular, if the theory undergoes a phase transition, the corrections are generally expected to be non-perturbative, and therefore it is not always possible to apply lightcone quantization without having already solved the theory.\footnote{In some cases, though non-perturbative, the corrections simply amount to a change in the bare parameters.}

The goal of this work is to point out a practical approach to get the best of both worlds.  Specifically, we want to obtain approximate volume independence and thus reliable results for correlation functions, while still accommodating a phase transition.  The idea is to take the standard equal-time quantization Hamiltonian truncation in finite volume, but consider states with a finite boost, or with a finite momentum.  We will show that even with moderate momenta one can already achieve volume independence over a large range of couplings.

In order to understand the effect of boosts, we first study one of the classic systems considered in the original Truncated Conformal Space Approach (TCSA) \cite{Yurov:1991my, Fonseca:2001dc}:  the Ising model deformed by thermal and magnetic field operators.  This system is an ideal toy model as it conveniently has both integrable and non-integrable limits, a phase transition in the thermal deformation, as well as universal features that are generic to any case of a CFT deformed by relevant operators.
We first begin with the two integrable deformations - deformation with only the thermal operator and deformation with only the magnetic operator.  In both cases, we study the volume-dependence in standard TCSA (at zero momentum) of the two-point function of the trace of the stress tensor, also known as the $c$-function.  We show that even with $\CO(30\,000)$ basis states, the $c$-function is strongly dependent on the volume outside of at most a narrow range of couplings.  In the case of the thermal deformation by the $\varepsilon$ operator, the dependence is so strong that no reliable determination of the infinite volume limit is possible.  In the case of the magnetic deformation $\sigma$, the contributions of some bound states also have a considerable variation with volume. 

We then show in section \ref{sec:BoostTCSA} that with even moderate boosts, the volume-dependence of the $c$-function can be drastically reduced, yielding stable results which agree well with integrability. 
Next, we consider generic non-integrable cases, both on the symmetry preserving and on the symmetry broken side. In contrast to standard TCSA, in all cases observables in the boosted frame display volume independence over a range of couplings, producing a universal volume-independent extrapolation curve.   For small explicit breaking (i.e. for a small magnetic field), we also examine the spectrum of the Hamiltonian, which is expected to undergo a phase transition as a function of the thermal deformation parameter.   This phase transition is more clearly visible at finite momentum.  Indeed, the spectra on each side of the phase transition appear nearly independent of volume while at the same time being very distinct from each other, as shown in Fig.~\ref{fig:spectrum-eta-pm4}.

We conclude in section \ref{sec:Discussion} with a brief examination of the connection between the boosted Hamiltonian and the lightcone quantization approach.  In particular, in the numerical spectrum of the lightcone Hamiltonian $P_+$, one can see the appearance of a clear low-energy sector accumulating near $P_+=0$ as the boost is increased.  This sector suggests that one can, with a proper definition of an effective Hamiltonian, $P_+^{\rm eff}$, restrict the theory entirely to a chiral basis which has $P_+=0$ before turning on the deformations.  Such a formulation will not have any volume dependence.
\section{TCSA in the Rest Frame}\label{sec:RestFrameTCSA}

In this paper, we consider the example of the thermal deformation and magnetic deformation of the 2d Ising CFT, one of the classic systems studied by TCSA. In TCSA, we put a CFT Hamiltonian $H_0$ on a cylinder of radius $R$, and deform it by CFT relevant operators to reach the target QFT in the IR. The Hamiltonian we are interested in is given by 
\begin{equation}\label{eq:Hamiltonian}
	H=H_{0}+\frac{1}{2\pi}\int_{0}^{2 \pi R} d x \,\big(  m\varepsilon(x) + g\sigma(x) \big) \, .
\end{equation}
with
\begin{equation}\label{eq:IsingCFTH}
	H_{0}=\frac{1}{2 \pi} \int_0^{2\pi R}dx(\psi \bar{\partial} \psi+\bar{\psi} \partial \bar{\psi}).
\end{equation}
The infinite volume limit is reached via taking $R\rightarrow \infty$, or equivalently $m, g \rightarrow \infty$ in units of $R=1$.
In the continuum limit, the theory depends only on the dimensionless parameter 
\begin{equation}
\eta = \frac{m}{h^{8/15}},\qquad \textrm{with } g=2\pi h.
\end{equation}
The Ising CFT with the Hamiltonian given in \eqref{eq:IsingCFTH} is described by a massless free fermion, and the $\varepsilon$ deformation is just the fermion mass term. So the $g=0$ case is integrable and can be used to check the numerical result. The $m=0$ case is also well understood because the $\sigma$ deformation is integrable. In this section, we study these two cases using TCSA in the rest frame, as baselines for improvement in the next section. 

 We provide the details of TCSA for the Ising model in appendix \ref{app:TCSA}. The idea of TCSA is simply to express IR physics in terms of the UV CFT states. 
In practice, in order to be able to perform the computation on a computer, we truncate the basis such that it only includes states that satisfy $E\leqslant \Lambda$, where  $E$ here is the energy of a CFT basis state (i.e., its $H_0$ eigenvalue) and $\Lambda$ is an energy cutoff. In this paper, the results are mostly obtained with $\Lambda\simeq 40$ (in units where $R=1$). For reference, for the rest frame TCSA that we are considering in this section, there are 26223 states for $\Lambda=40$, including both the Neveu-Schwarz sector and the Ramond sector.

 We will be interested in the Zamolodchikov $c$-function, which can be computed by integrating (or summing over) the spectral density of the trace of the stress-energy tensor $\Theta$. Specifically,
the trace of the stress-energy tensor for the Hamiltonian \eqref{eq:Hamiltonian} is given by 
\begin{equation}
	\Theta=\frac{1}{2\pi}\left(\left(2-\Delta_{\varepsilon}\right) m \varepsilon+\left(2-\Delta_{\sigma}\right) g \sigma\right),  
\end{equation}
where $\Delta_\varepsilon=1$ and $\Delta_\sigma=1/8$. In TCSA, the spectral density is computed by 
\begin{equation}\label{eq:RestFrameSD}
	\rho_{\Theta}\left(\mu^2\right)=\sum_{i\ge 1}\left|\left\langle\Omega|\Theta| E_{i}\right\rangle\right|^{2}\delta(\mu^2-\mu_i^2), \quad \textrm{with }\mu_i\equiv E_i-E_0\end{equation}
where $|\Omega\rangle$ is the vacuum state and  $|E_i\rangle$ is the $i+1$-th eigenstate ($i\ge 1$) of the Hamiltonian \eqref{eq:Hamiltonian} with eigenvalue $E_i$ (in this convention, $|\Omega\rangle\equiv|E_0\rangle $)\footnote{Here, we are assuming that the states are orthogonal and normalized to have norms $\langle E_i|E_i\rangle=4\pi\mu_i $ for $i\ge 1$ and $\langle E_0|E_0\rangle=1$.}. 
Using $\rho_\Theta$, we can then compute  the $c$-function via 
\begin{equation}\label{eq:cFun}
	c(\mu)= 12 \pi \int_0^{\mu^2} d \mu'^2 \frac{\rho_{\Theta}(\mu'^2)}{\mu'^4}.
\end{equation}
In this section, we will focus on computing the mass gaps ($m_\textrm{gap}=E_1-E_0$) and the $c$-functions of the $\sigma$ deformation and the $\varepsilon$ deformation of the Ising model. 

\subsection{Thermal deformation of the Ising model}

\begin{figure}[htbp]
\centering
\includegraphics[width=0.45\linewidth]{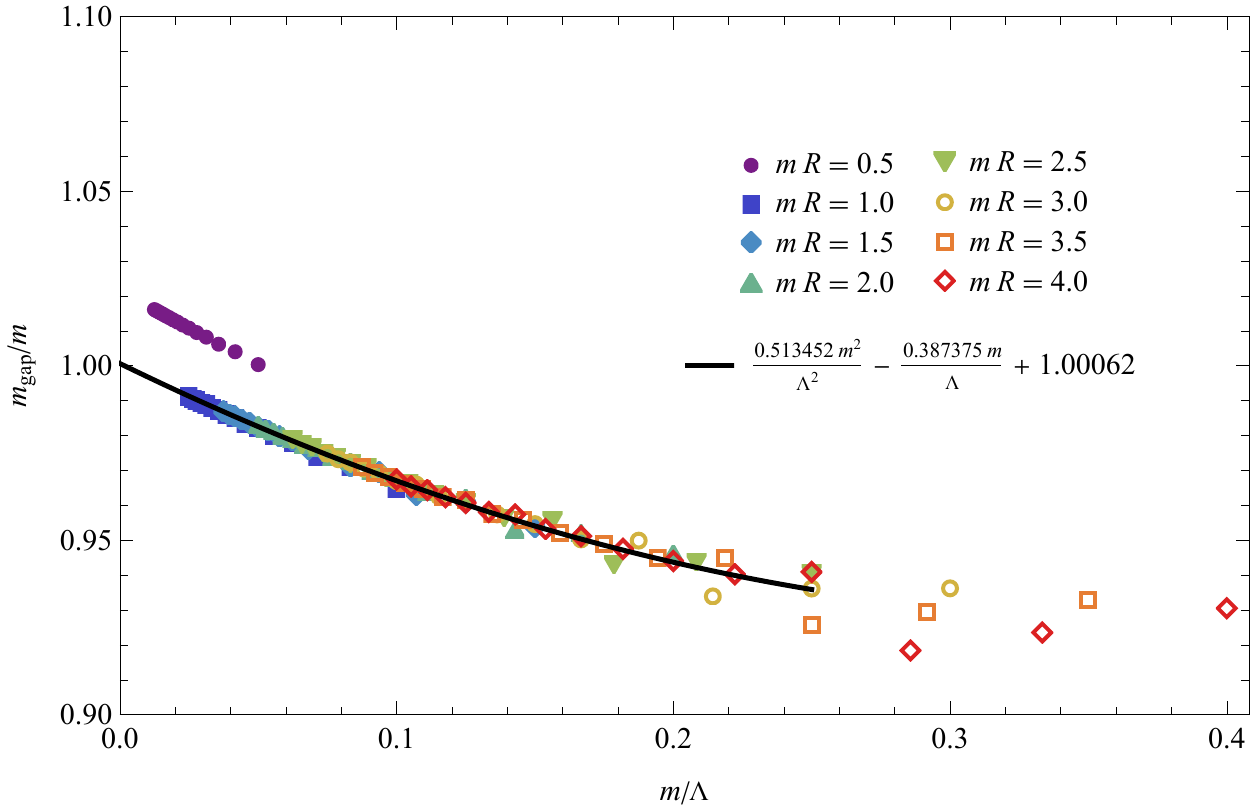}
\includegraphics[width=0.45\linewidth]{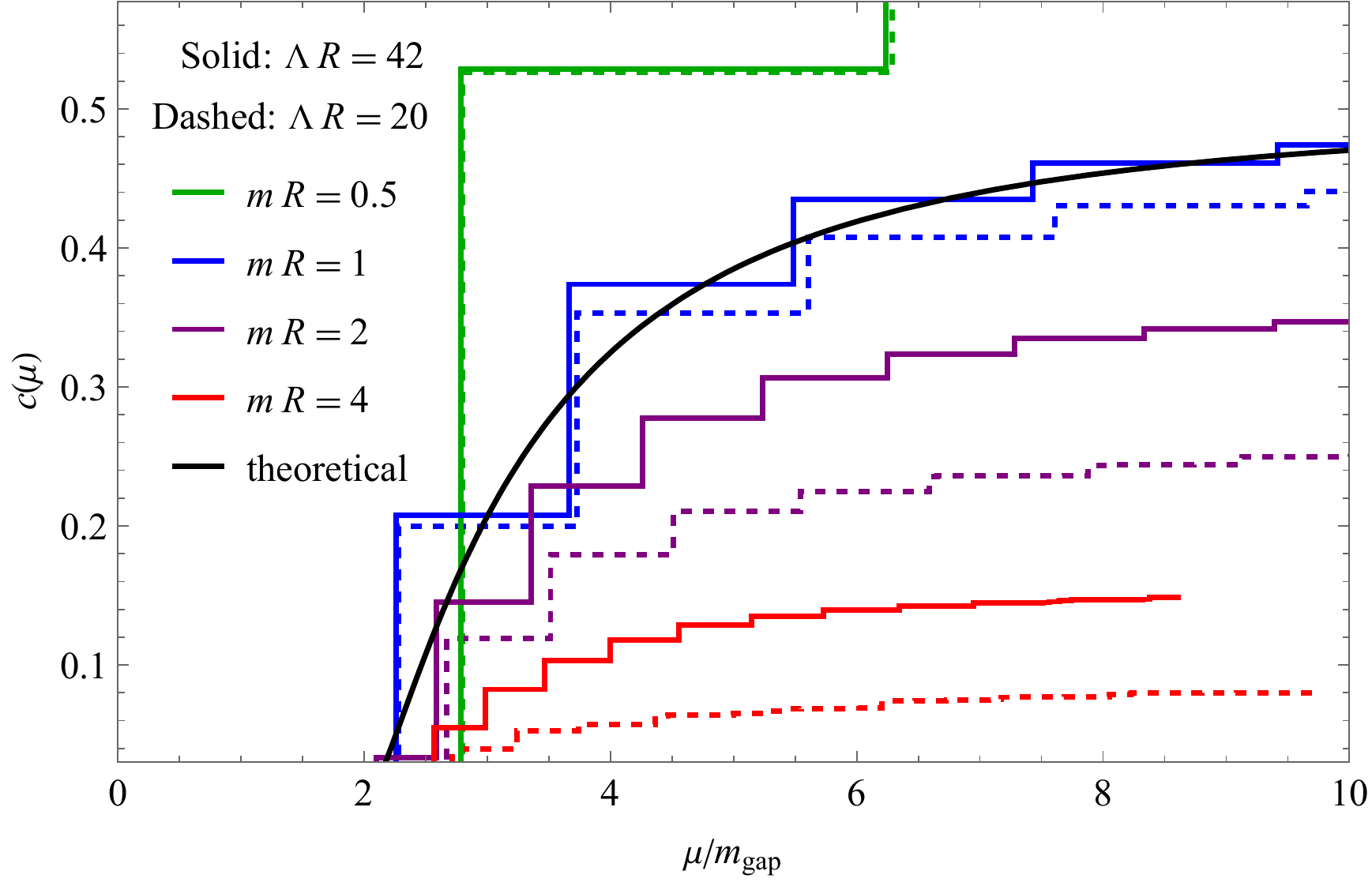}
\caption{
	\label{fig:epsilon-rest}
		{\bf Left panel:} 
	The mass gap $m_{\rm gap}$ measurement in the rest frame.
	The plot shows the dependence of $m_{\rm gap}$ as a function of $m/\Lambda$. For $m R\geqslant1.0$, the results at different $\Lambda$ and $m$ form a universal behavior. $m R =0.5$ sits outside of the universal fit, indicating that the finite-volume effect is significant.
	{\bf Right panel:} Measurement of the $c$-function in the rest frame. Because truncation always renders a discrete spectrum, the resulting $c$-function is a series of steps. 
	For $m R=1$, the truncation result roughly agrees with the theoretical result (black line) given in equation \eqref{eq:cfunctionepsilon}. The truncation result for $m R=0.5$ deviates from the theoretical result because of finite-volume effects. For $m R=2$ and $mR= 4$, the deviation from the theoretical result is due to truncation error, as can be seen by comparing the results at $\Lambda R=20$ and $\Lambda R = 42$. 
}
\end{figure}
For the $\varepsilon$ deformation, 
to minimize the finite-volume corrections, we need to take $m$ large. However, for the truncation to be precise, we need $\Lambda \gtrsim m$, which causes the number of states required to grow exponentially. This is a numerical manifestation of Haag's theorem in finite volume. Whether a good truncation result is reached, depends on the existence of a Goldilocks regime where $m$ is sufficiently large to suppress the finite volume correction while the truncation error is still under control. The numerical result of measuring the mass gap $m_\textrm{gap}=E_1-E_0$ using the first two eigenvalues of the Hamiltonian is shown in the left panel of Figure \ref{fig:epsilon-rest}. The nice agreement between various $m$ can be used to obtain an extrapolation to the infinite volume, infinite truncation limit that agrees with the known result at better than $0.1\%$ accuracy. It seems to suggest that the truncation is accurate for a wide range of $m$.

However, if we take a step further to measure the Zamolodchikov $c$-function\footnote{In this case, the $c$-function is simply given by 
\begin{equation}\label{eq:cfunctionepsilon}
	c(\mu)=\frac{1}{2} \left(\frac{\mu ^2-4m^2}{\mu ^2}\right)^{3/2}.
\end{equation}
}, as shown in the right panel of Figure \ref{fig:epsilon-rest}, we see that we do not have a stable measurement, consistent over a range of $m$. Essentially, we get a numerical result that roughly matches the correct answer only if we tune close to $mR=1$, and without a priori knowledge of the exact answer, we would have no way to know this. In a generic situation where we do not know the answer ahead of time, to extract the $c$-function from truncation results it is necessary for them to be stable over a range of $m$.  Note that, as shown in the right panel of Figure \ref{fig:epsilon-rest}, at $mR =1$, the $c$-function has roughly converged as a function of $\Lambda$ by $\Lambda=20$, but for larger values of $m R$ it has not and is moving significantly.  Thus $m R=1$ is at a kind of sweet spot for the value of $m$ in that finite volume corrections are suppressed, but truncation effects have not yet started to grow significantly.  We expect that the main truncation effects at large $m$ are due to the orthogonality catastrophe. We will see that at both $m R < 1$, where finite-volume effects dominate, and $mR >1$, where truncation effects dominate, going to a boosted frame drastically improves the results for the $c$-function.

\subsection{Magnetic deformation of the Ising model}\label{sec:sigmaDeformation}

In this subsection, we study the $\sigma$ deformation of the Ising model, and will find that the situation is similar to what we found for the $\varepsilon$ deformation in the previous subsection. That is, while one can obtain a precise measurement of the mass gap from TCSA in the rest frame, the measurement of the $c$-function suffers from much more volume dependence which makes it more challenging to robustly extract the infinite volume behavior.

As mentioned above, the magnetic deformation of the Ising model by the $\sigma$ operator is also integrable. The spectrum consists of eight bound states whose masses $m_i$ are related by an $E_8$ symmetry. Each bound state has a finite contribution to the Zamolodchikov $c$-function. These are all known analytically, see \cite{Delfino:1995zk} for the derivation. We present the lowest spectrum and their contributions to the $c$-function that are relevant to the discussion in this paper in the following table for later reference:

\begin{table}[h]
\begin{center}
\begin{tabular}{|c|c|c|}
\hline  \text{index}  &  \ensuremath{m_{i}}  &  \ensuremath{c_{i}} \\
\hline  $i=1$  &  \ensuremath{\mathcal{C}h^{8/15}} = (4.40490858...) \ensuremath{h^{8/15}}  &  0.472038282 \\
\hline  $i=2$  &  2 \ensuremath{m_{1}} \ensuremath{\cos\frac{\pi}{5}} =(1.6180339887 \ensuremath{\ldots}) \ensuremath{m_{1}}  &  0.019231268 \\
\hline  $i=3$  &  2 \ensuremath{m_{1}} \ensuremath{\cos\frac{\pi}{30}} =(1.9890437907 \ensuremath{\ldots}) \ensuremath{m_{1}}  &  0.002557246 \\
\hline  $i=4$  &  2 \ensuremath{m_{1}} \ensuremath{\cos\frac{7\pi}{30}} =(2.4048671724 \ensuremath{\ldots}) \ensuremath{m_{1}}  &  0.000700348 
\\\hline \end{tabular}
\end{center}
\caption{Masses and contributions to the $c$-function from the first several bound states of the $\sigma$ deformation of the Ising model \cite{Delfino:1995zk}.}
\label{table:E8}
\end{table}
\noindent The constant $\mathcal{C}$ in the above table is given by 
\begin{equation}\label{eq:epsilonGap}
\mathcal{C}=\frac{4 \sin \frac{\pi}{5} \Gamma\left(\frac{1}{5}\right)}{\Gamma\left(\frac{2}{3}\right) \Gamma\left(\frac{8}{15}\right)}\left(\frac{4 \pi^{2} \Gamma\left(\frac{3}{4}\right) \Gamma^{2}\left(\frac{13}{16}\right)}{\Gamma\left(\frac{1}{4}\right) \Gamma^{2}\left(\frac{3}{16}\right)}\right)^{4 / 15}
\, .
\end{equation}
Our task in this subsection is to compute these observables in TCSA in the rest frame and compare them to the theoretical results. Note that the first three bound states in the above table are below the two-particle continuum of the first bound state, and these will be our focus in this subsection.

\begin{figure}[t]
\begin{center}
\qquad\qquad\qquad\includegraphics[width=0.7\textwidth]{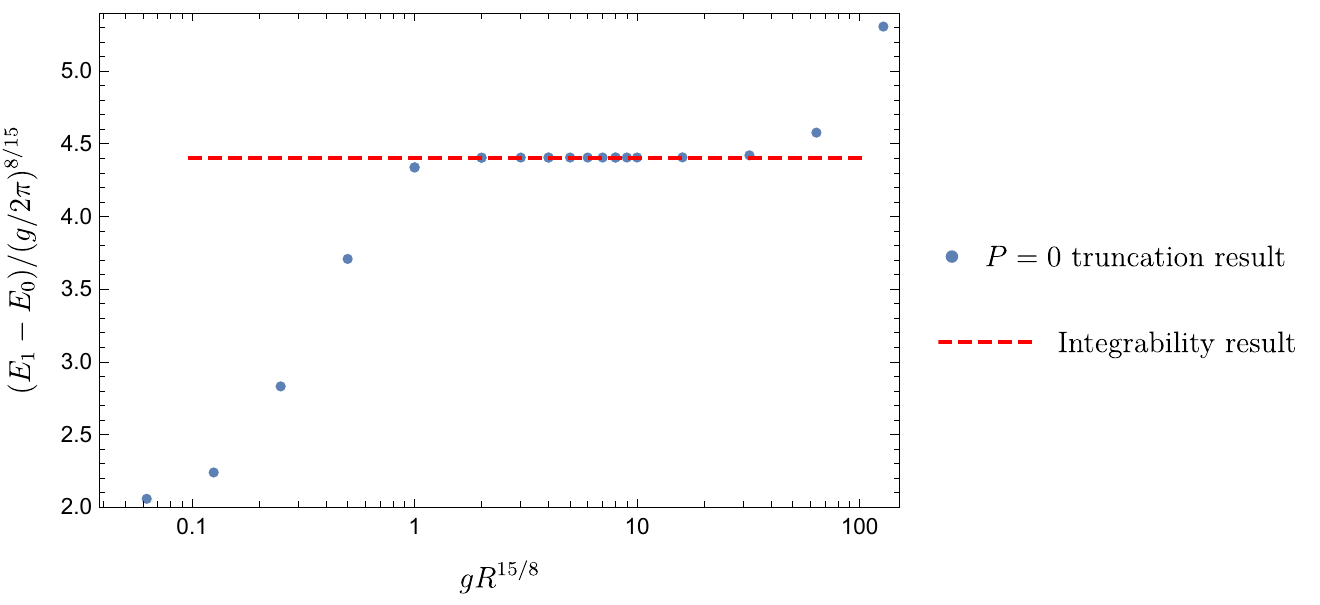}

\caption{The relationship between the mass gap and the coupling constant $g=2\pi h$ can be obtained using integrability, which gives $m_1/h^{8/15}=4.40490857...$. From truncation, we can determine this number by finding the range in $g$ where $(E_1-E_0)/h^{8/15}$ is constant, such as $1\lesssim g R^{15/8}\lesssim 32$ in the above plot. Simply picking $gR^{15/8}=5$, we have $(E_1-E_0)/h^{8/15}=4.404921$, which is in good agreement  with the theoretical result.}
\label{fig:sigmaDeformationMassGap}
\end{center}
\end{figure}

\begin{figure}[h!]
\begin{center}
\includegraphics[width=0.48\textwidth]{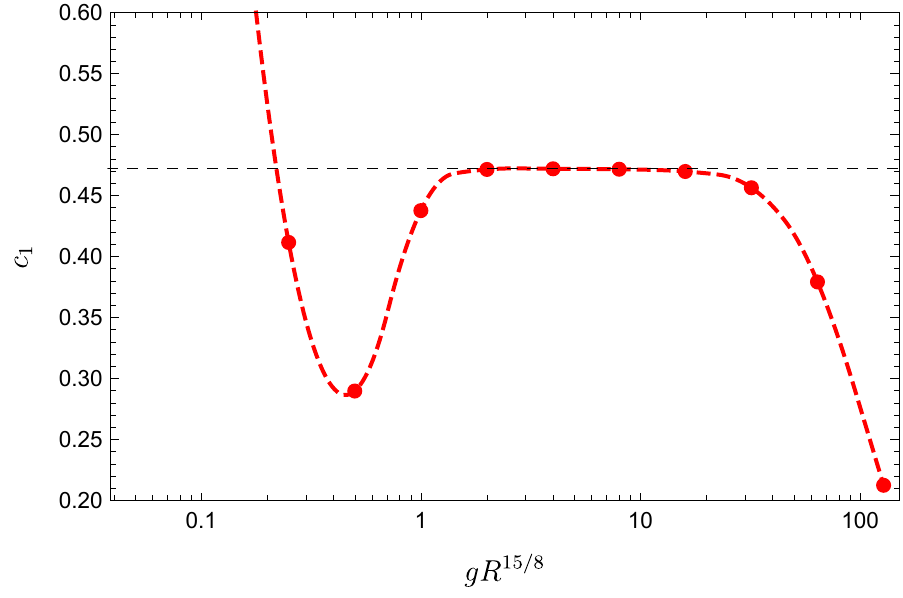}
\quad
\includegraphics[width=0.48\textwidth]{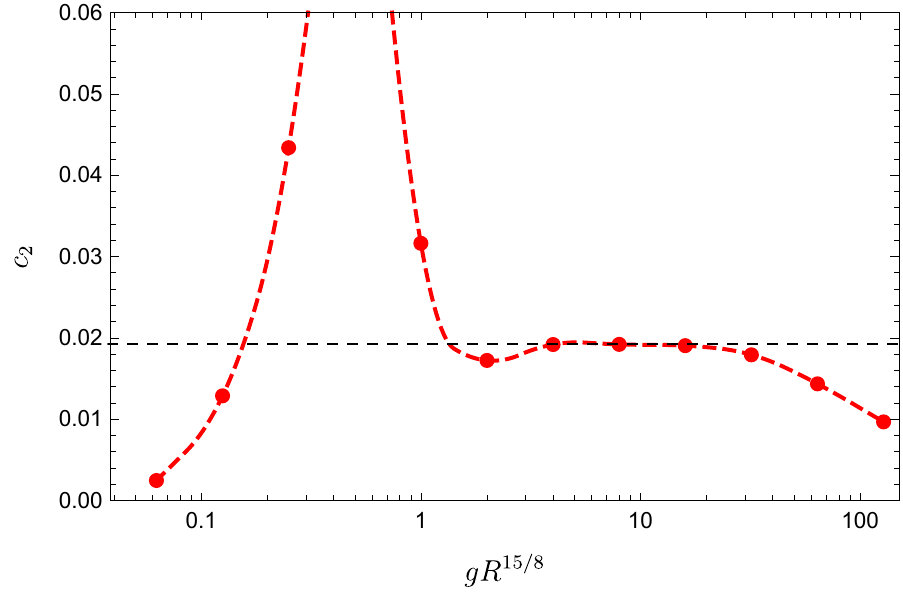}\\
\vspace{0.3cm}
\includegraphics[width=0.48\textwidth]{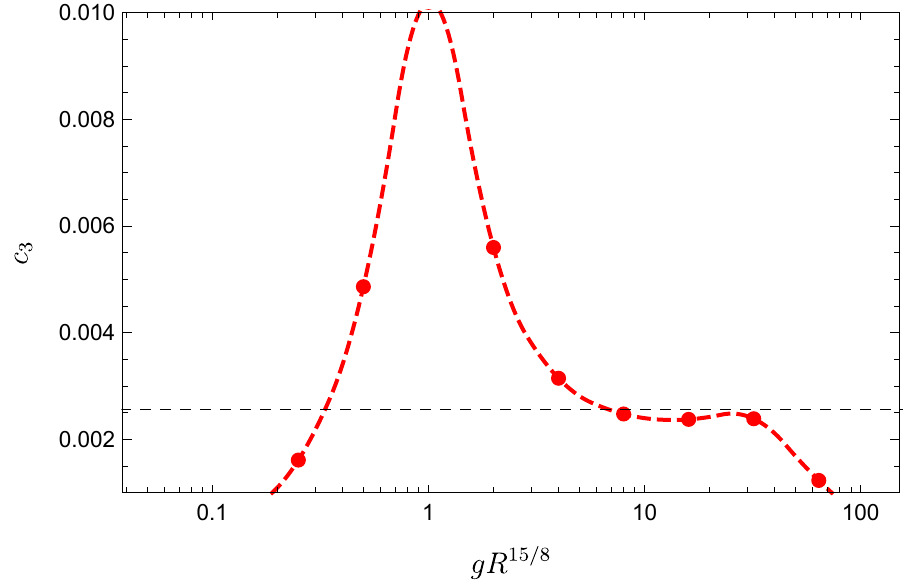}
\quad
\includegraphics[width=0.465\textwidth]{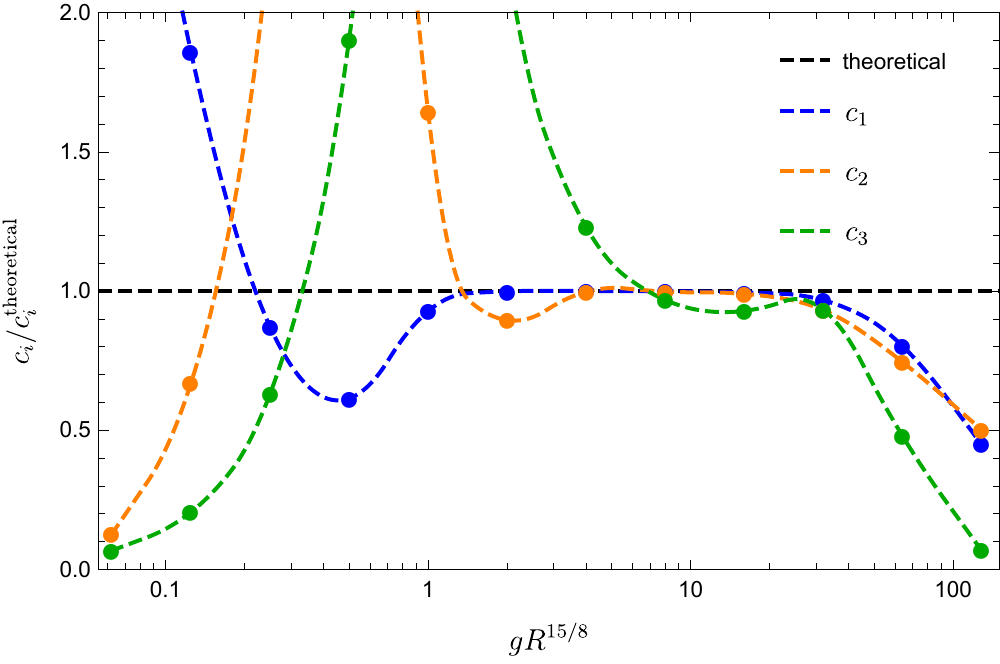}
\caption{Contributions to $c$-function from the first three $E_8$ states computed using TCSA in the rest frame, with $\Lambda R=40$. The last plot is an aggregation of the first three plots. The black dashed lines are the theoretical results from integrability given in Table  \ref{table:E8} for comparison.}
\label{fig:sigmaDeformationDeltaCiP0}
\end{center}
\end{figure}

The magnetic deformation of the Ising model is one of the first several models studied using TCSA in the literature \cite{Yurov:1991my}. The computation done in this subsection is exactly the same as in \cite{Yurov:1991my}, except that we go to higher truncation (we use $\Lambda=40$ with 26223 states compared to $\Lambda=11$ with 39 states in \cite{Yurov:1991my}).
The result for the mass gap $m_\textrm{gap}=E_1-E_0$ is given in Figure \ref{fig:sigmaDeformationMassGap}. One can see that in the range $1\lesssim g R^{15/8}\lesssim 32$, the mass gap is basically given by\footnote{Here we simply pick the value measured at $g R^{15/8}=5$, since the result is stable over a large range of $g$.} $m_1=\mathcal{C}'\left(\frac{g}{2\pi}\right)^{8/15}$ with  $\mathcal{C}'\simeq 4.404921$, which is very close to the integrability result  (\ref{eq:epsilonGap}). In fact, the dominant finite volume corrections to the mass gap are the so-called ``$\mu$'' terms, from the production of virtual particles going `around the world' \cite{Klassen:1990ub}. We will see that this leading correction is suppressed in a boosted frame. 

However, if we measure the contributions to the $c$-function from the first three bound states,  shown in Figure \ref{fig:sigmaDeformationDeltaCiP0}, we see that the range where the truncation result agrees with the theoretical result shrinks as we go from $c_1$ to $c_3$. The result is especially bad for $c_3$, where it only roughly agrees with the theoretical result for a small range of $g$. Therefore, the situation here is similar to the $\varepsilon$ deformation case that we discussed in the last subsection, that is, the measurement of the mass gap is accurate in TCSA in the rest frame, but not so for other more complicated observables.

\section{TCSA in Boosted Frames}
\label{sec:BoostTCSA}

In the last section, we have studied TCSA in the rest frame. Although the mass gap measurement is accurate in a range of volumes, the $c$-function suffers from strong volume dependence, and no reliable measurement can be extracted, especially for the $\varepsilon$ deformation. In this section, we aim to improve the truncation results by 
going to boosted frames.

In a boosted frame with momentum $p$, the TCSA basis consists of CFT states with  momentum $p$, but otherwise the Hamiltonian truncation calculation follows the traditional rest frame computation. The Hilbert space is truncated by only including states that satisfy $E^2-p^2\le\Lambda^2$. In Table \ref{table:numberOfStates}, we provide a comparison of the basis sizes for different values of $\Lambda$ and $p$. One can see that for the cases considered in this paper, the basis size does not depend on $p$ much, and boosting actually decreases the basis size slightly\footnote{This is not the case if we continue to increase $p$. For example, for $pR=14$ and $\Lambda R=40$, there are 30763 states. The point is that the dependence of the basis size on $p$ is weak in the regime that we are considering, compared to exponential dependence on $\Lambda$.}.   For more details, see appendix \ref{app:TCSA}.  

\begin{table}
\begin{center}
\begin{tabular}{|c|c|c|c|c|c|c|c|}
\hline 
 & $p R=0$ & $p R=2$ & $pR=4$ & $pR=6$ & $pR=8$ & $pR=10$ & $pR=12$\tabularnewline
\hline 
$\Lambda R=40$ & 26223 & 26067 & 25841 & 25211 & 24593 & 25810 & 24764\tabularnewline
\hline 
$\Lambda R=42$ & 37104 & 36998 & 36501 & 35956 & 34862 & 36906 & 35222\tabularnewline
\hline 
\end{tabular}
\end{center}
\caption{Number of states for different values of $p$ and $\Lambda$ that are used in this paper.}
\label{table:numberOfStates}
\end{table}

 One subtlety about doing TCSA in a boosted frame is that the vacuum is not included in the boosted sector of the Hilbert space, which has two consequences.
First, we still need to use the rest frame vacuum state $|\Omega\rangle$ when computing vacuum expectation values or spectral densities. 
Second, the effective vacuum energy seen by the boosted frame spectrum also needs to be measured from the rest frame data. Practically, we have found that the best procedure for determining the vacuum energy in a boosted frame is to require that the mass gaps measured in a boosted frame and the rest frame are the same\footnote{Interestingly, this procedure works better than directly using the vacuum energy measured in the rest frame as the boosted frame vacuum energy. This is likely due to the breaking of Lorentz invariance due to finite volume effects, leading the vacuum energy to have a weak dependence on the inertial frame.}. To be concrete, let us denote the eigenstates of the Hamiltonian \eqref{eq:Hamiltonian} in a boosted frame with momentum $p$ by $|E_i^p\rangle$ (with $i\ge 1$, since the first eigenstate is a single-particle state instead of the vacuum), where $E_i^p$s are the corresponding eigenvalues.  We compute the vacuum energy in this boosted frame by the following formula\footnote{For the $\varepsilon$ deformation case, we restrict to the NS sector, and $m_{\rm gap}$ means the two-particle threshold, since in this case the first non-vacuum state is a two-particle state.} 
\begin{equation}\label{eq:BoostVacuumE}
E_{\rm vac}^p = E_{1}^p - \sqrt{ m_{\rm gap}^2 + p^2 } , 
\end{equation}
where $E_1^p$ is the first eigenvalue of the Hamiltonian in this boosted frame, and $m_{\rm gap}$ is measured in the rest frame as we did in the last section. Then, the invariant mass of a boosted frame eigenstate with eigenvalue $E_i^p$ is given by
\begin{equation}\label{eq:BoostSpectrum}
\mu_i = \sqrt{ ( E_i^p - E_{\rm vac}^p )^2 - p^2 }.
\end{equation}
The spectral density of the trace of the stress-energy tensor $\Theta$ is given by 
\begin{equation}
	\rho_{\Theta}\left(\mu^2\right)=\sum_{i\ge1} \left|\left\langle\Omega|\Theta| E_i^p\right\rangle\right|^{2}\delta(\mu^2-\mu_i^2), 
\end{equation}
where $|\Omega\rangle$ is the rest frame vacuum and the states are normalized to have norms $\langle E_i^p| E_j^p\rangle=4\pi\sqrt{\mu_i^2+p^2}\delta_{ij}$. Note that this is very similar to the rest frame formula given in \eqref{eq:RestFrameSD} except that the normalization of the states is different. The $c$-function in the boosted frame is again given by equation \eqref{eq:cFun}.

The procedure outlined here is used to compute the $c$-function from the boosted frame TCSA data in the rest of this section.

 \subsection{Obtaining accurate \texorpdfstring{$c$}{c}-functions from TCSA  in boosted frames}
In this subsection, we use the procedure outlined above to compute the $c$-functions for the two integrable deformations of the Ising model, and show that the results obtained this way are much more accurate than those given in section \ref{sec:RestFrameTCSA}. 
 
\begin{figure}[htbp]
\centering
\includegraphics[width=0.45\linewidth]{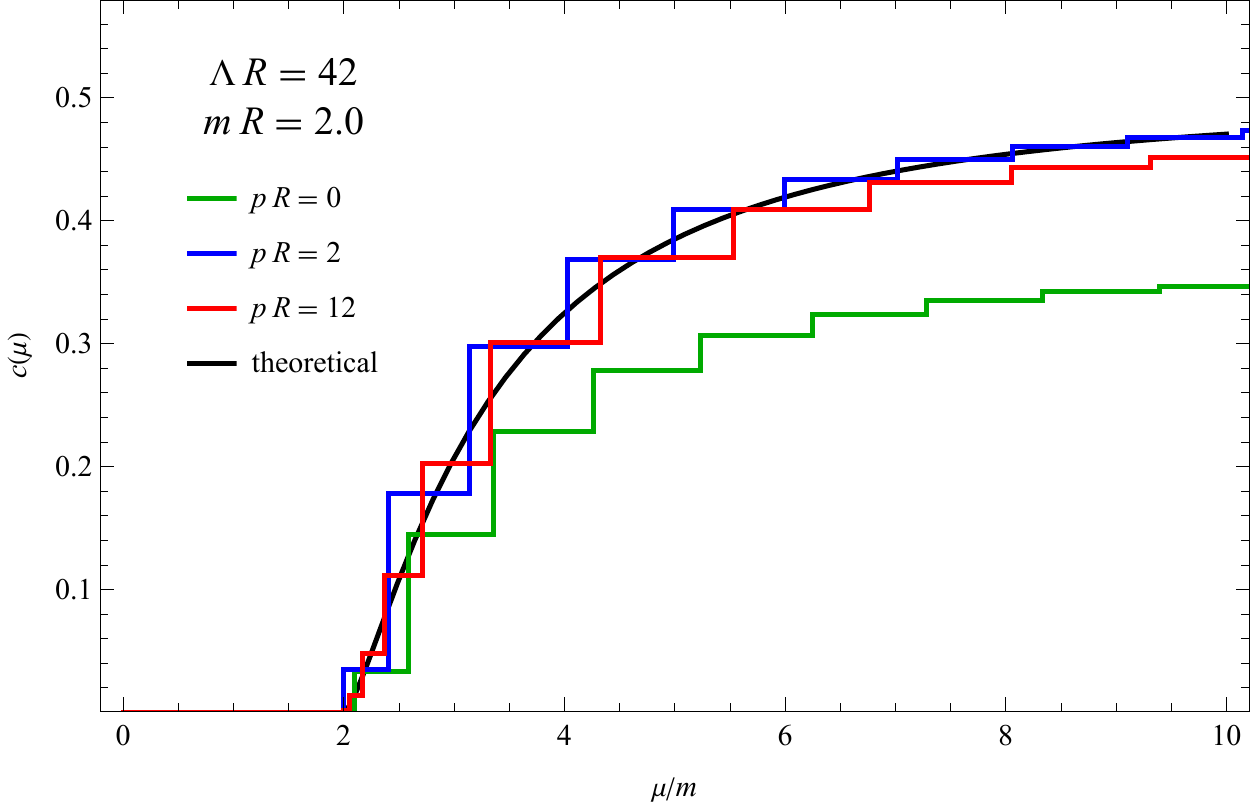}
\quad
\includegraphics[width=0.45\linewidth]{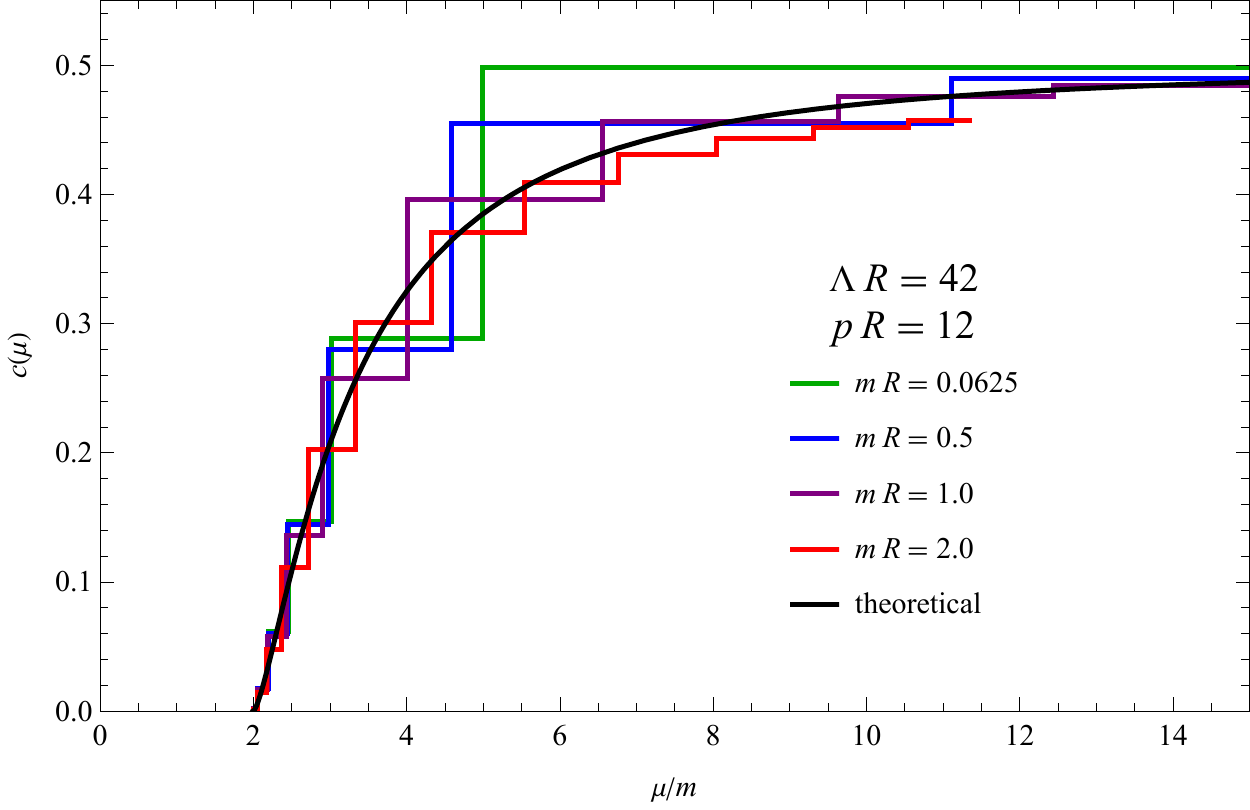}\\
\vspace{0.3cm}

\includegraphics[width=0.45\linewidth]{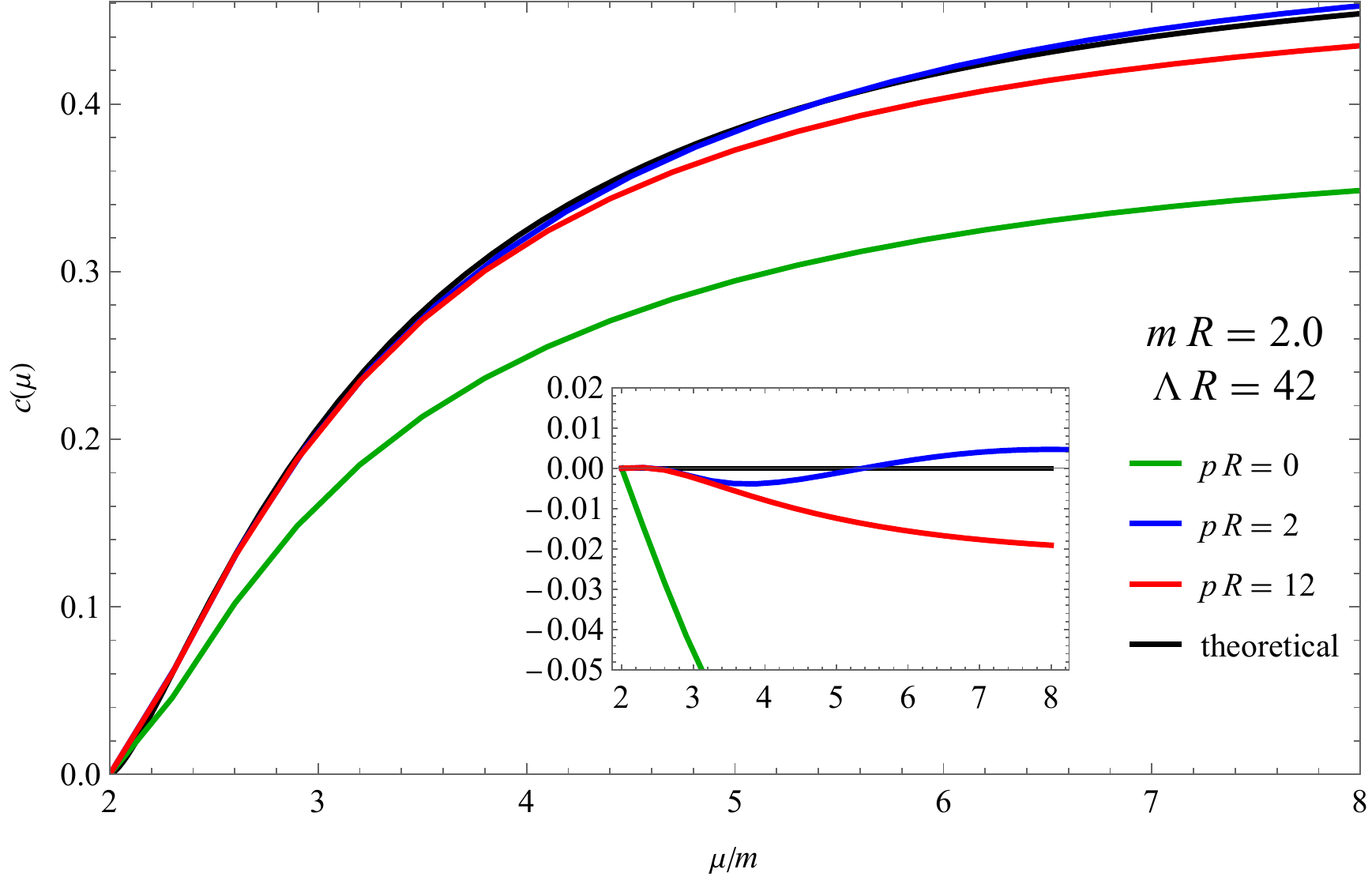}
\quad
\includegraphics[width=0.45\linewidth]{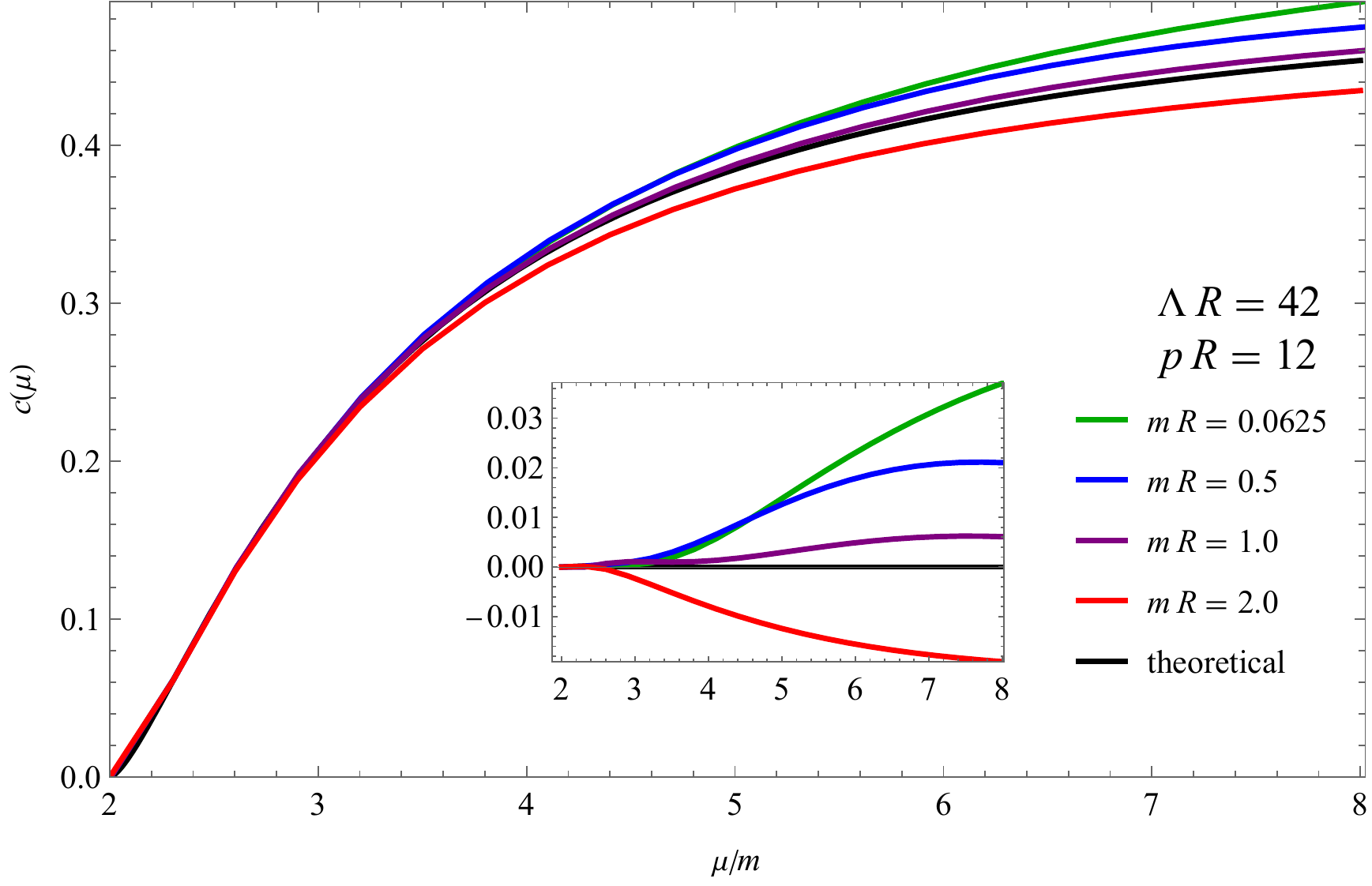}
\caption{
	\label{fig:epsilon-boost}
	Measurement of the $c$-function for the $\varepsilon$ deformation. {\bf Top left panel:} The $c$-functions for  $mR=2.0$ measured in the boosted frames with different values of $p$. 
	{\bf Top right panel:} The $c$-functions for   different values of $m$ measured in the boosted frame with large momentum $pR=12$. {\bf Bottom panels:} Smooth interpolations of the top panels, obtained by first taking a Pad\'e approximant of the momentum space time-ordered correlator $\< \Theta(p)\Theta(-p) \>$ in the variable $z$, where $p^2 = 16m^2/(z+z^{-1} +2)$, and then taking the imaginary part. 
	The details of this interpolation method are discussed in \cite{Chen:2021bmm}. The insets are the differences between the Pad\'e approximation results and the theoretical result. 
}
\end{figure}

\begin{figure}[htbp]
\centering
\includegraphics[width=0.45\linewidth]{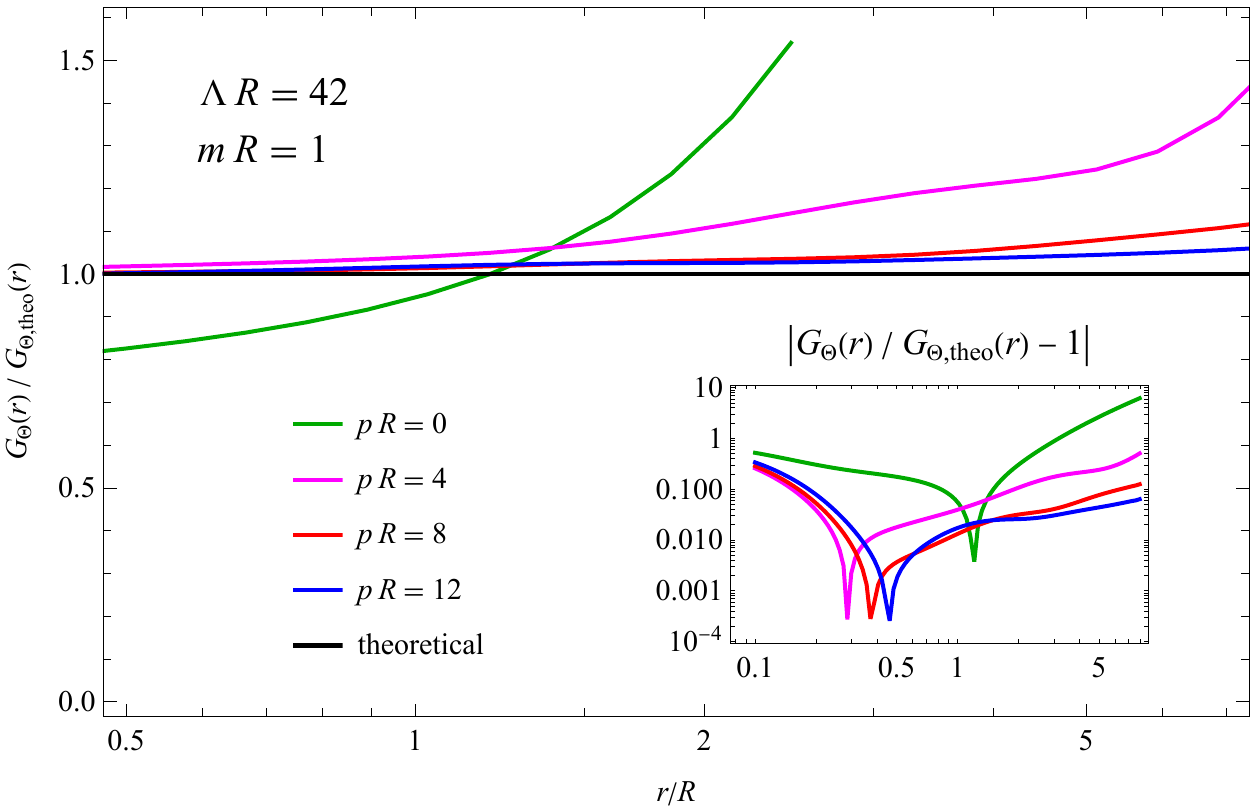}
\quad
\includegraphics[width=0.45\linewidth]{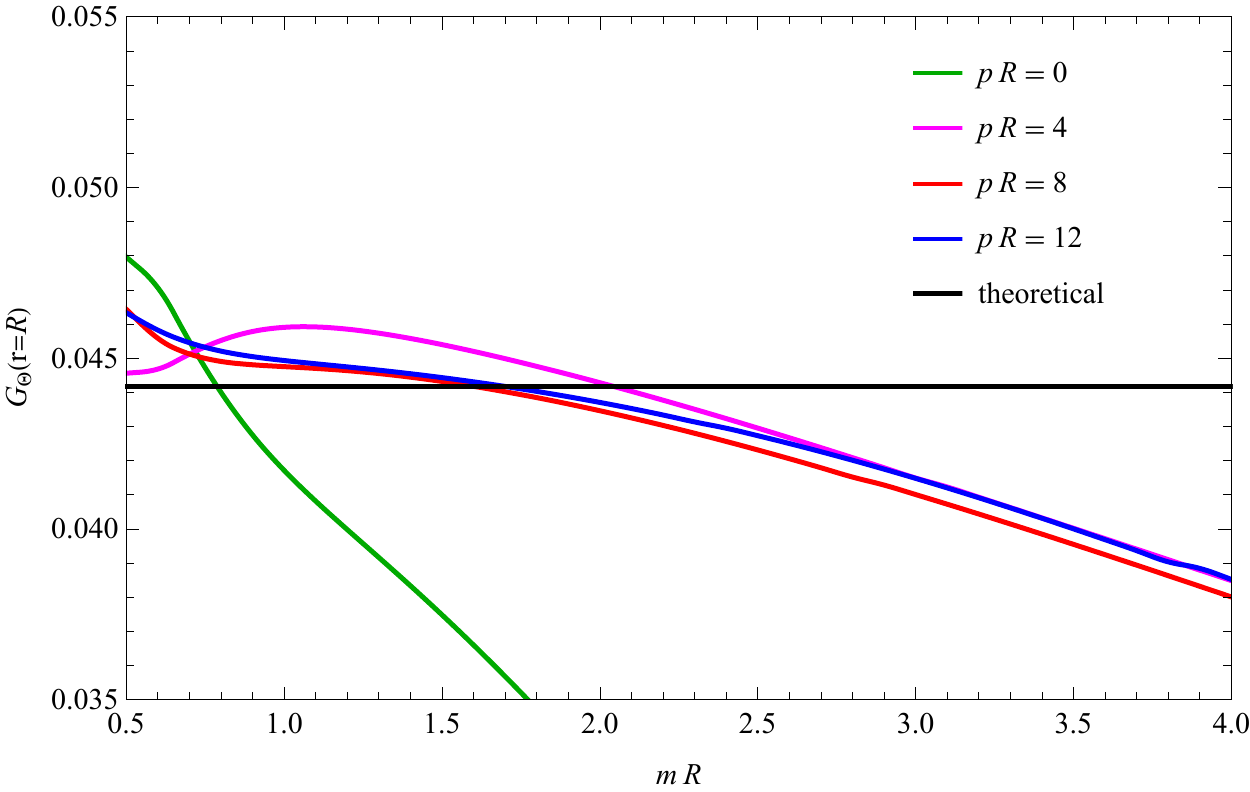} \\
\vspace{0.3cm}
\includegraphics[width=0.45\linewidth]{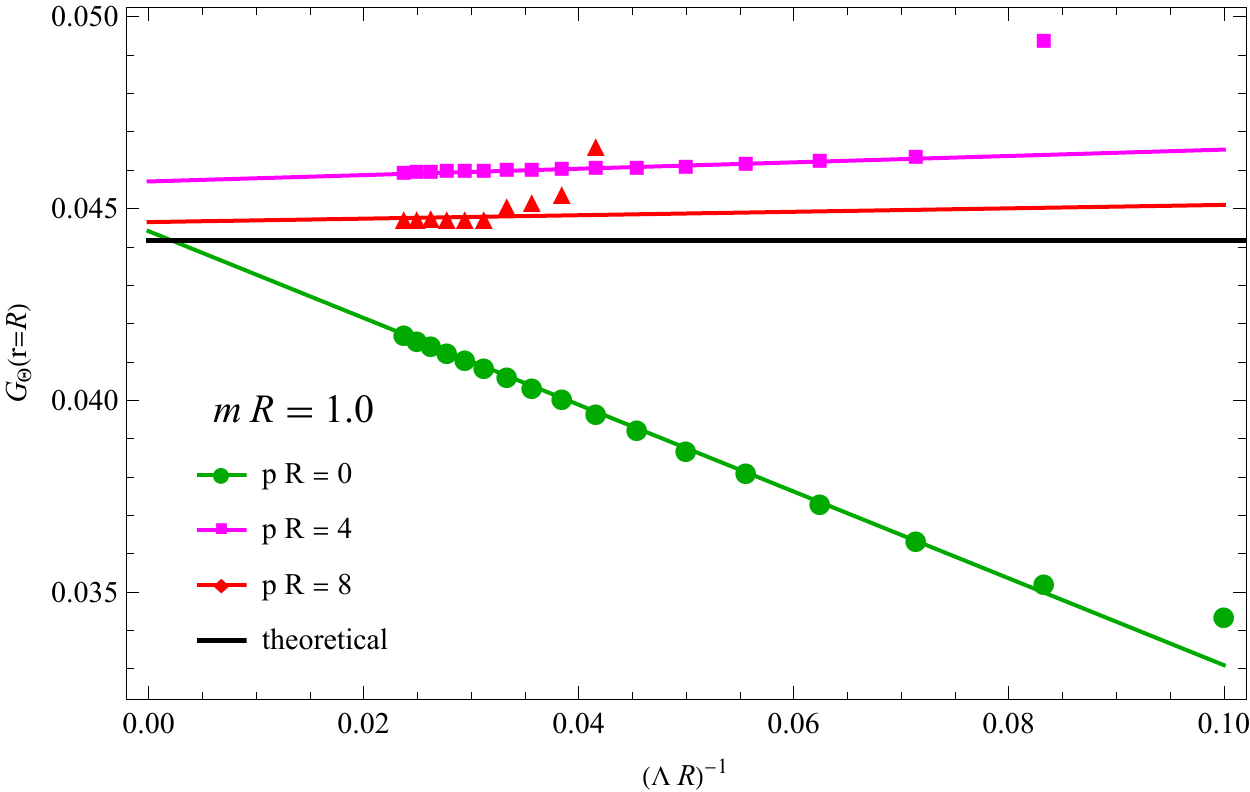}
\quad
\includegraphics[width=0.45\linewidth]{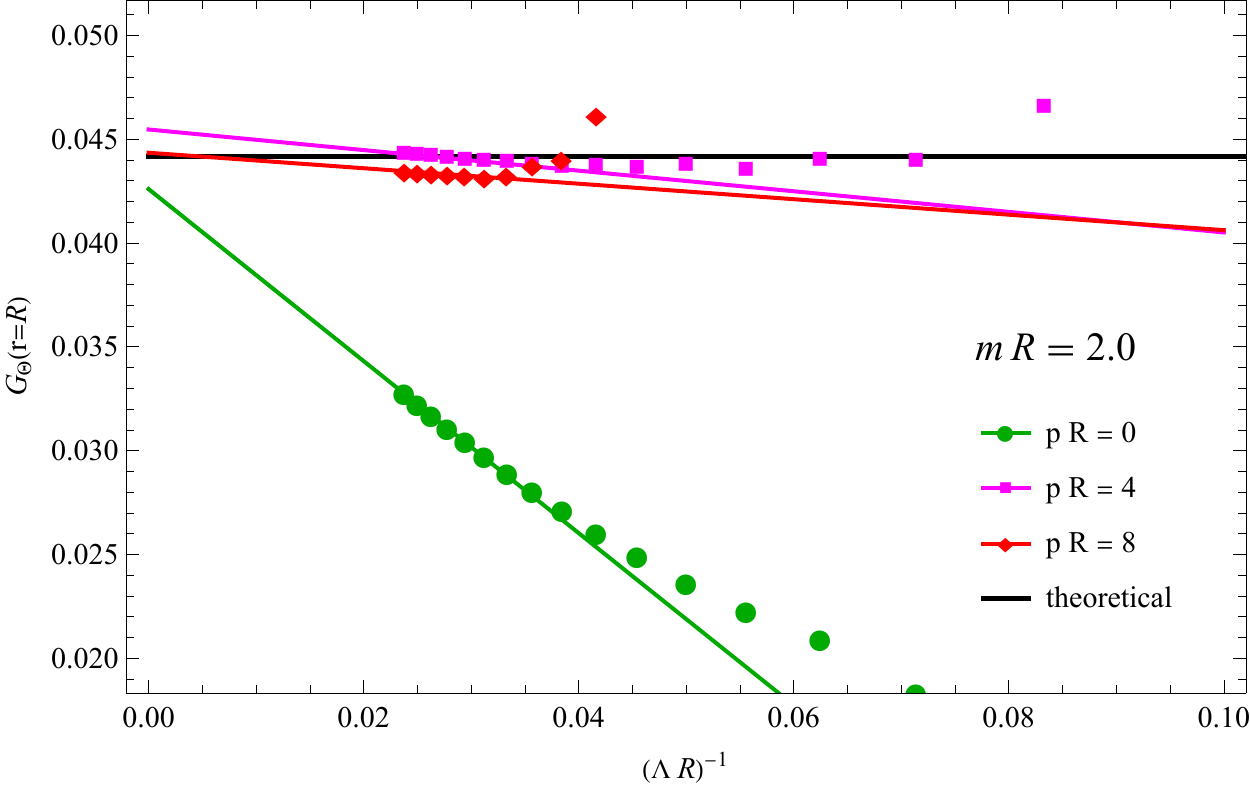} 
\caption{
	\label{fig:epsilon-spacial-correlator}
	Measurement of the spacial correlator $G_{\Theta}(r)$. {\bf Top Left:} Ratio of the truncation data $G_{\Theta}(r)$ to the theoretical result $G_{\Theta, {\rm theo}}(r)$, as a function of the space-like separation $r$. The rest frame result (green) never reaches a stable plateau at any values of $r$, while the boosted frame results are accurate for some range of $r$, and the range increases as we take higher boost. 
		{\bf Top Right:} $G_{\Theta}(r=R)$ at a fixed separation $r=R$, as a function of the system's volume $m R$. Again, the rest frame result (green) is never consistent for any range of $m$, while the boosted result stay within a few percent error for the range $1.0 < m R < 2.0$.
	{\bf Bottom Left:} Convergence of the truncation result for $G_{\Theta}(r=R)$ as a function of cutoff $\Lambda$, measured at $mR=1.0$. Compared to the rest frame result (green), the boosted frame result typically has a smaller error for a fixed $\Lambda$. Higher boost $pR=8$ result (red) has a better extrapolated result than the lower boost $pR=4$ result (magenta), while requiring a higher $\Lambda$ to reach the linear regime. 
	{\bf Bottom right:} Same as the bottom left plot, but at $mR=2.0$, which corresponds to a larger volume. The rest-frame result (green) is significantly worse than that of $mR=1.0$, while the boosted results remain accurate.
}
\end{figure}

For the $\varepsilon$ deformation, we take the universal fit of the mass gap in the left panel of  Figure \ref{fig:epsilon-rest} as the $m_{\rm gap}$ in equation \eqref{eq:BoostVacuumE} to obtain the vacuum energy in the boosted frame. Then, we use equation \eqref{eq:BoostSpectrum} to compute the center-of-mass energies of the spectrum, and the rest frame vacuum state to compute the spectral density of $\Theta$, which is then used to compute the $c$-function. We present the result of the  $c$-functions obtained this way in Figure \ref{fig:epsilon-boost}. From the top-left panel of Figure \ref{fig:epsilon-boost}, one can see that for a specific value of $m$, the $c$-functions obtained from the TCSA data in boosted frames with different values of $p$ agree much better with the theoretical result than the rest frame one. In the top-right panel of the Figure \ref{fig:epsilon-boost}, we show that the $c$-function is accurate for a significantly larger window of mass $m$ and energy scale $\mu$ in the boosted frame. Note that in the rest frame result shown in the right panel of Figure \ref{fig:epsilon-rest}, only the $m=1$ $c$-function is close to the theoretical result. Therefore, the boosted frame result shown in Figure \ref{fig:epsilon-boost} is a significant improvement compared to that of the rest frame. In the bottom panels of Figure \ref{fig:epsilon-boost}, we used the Pad\'e approximation trick of \cite{Chen:2021bmm} to get smooth interpolation of the plots in the top panels, which again confirmed that boosting gives better results.

It is also useful to look at the position space correlators $G_{\Theta}(r) \equiv R^4 \< \Theta(x) \Theta(0) \>$ for a space-like separation $|x| = r$, which can be obtained from the spectral density $\rho_{\Theta}$ as an integral
\begin{equation}
\begin{aligned}
G_{\Theta}(r) = 
R^4\int d\mu^2 \rho_{\Theta}(\mu) \int \frac{d^2p}{2\pi} \delta(p^2 - \mu^2) e^{ip\cdot x}
= R^4\int d\mu^2 \rho_{\Theta}(\mu) \frac{1}{2\pi} K_0(\mu r) \, .
\end{aligned}
\end{equation}
The advantage of this measurement is that it also smooths out the discrete truncation result for $\rho_{\Theta}$, so we can see the details of $\rho_{\Theta}$ hidden by the discrete jumps. The position space correlator results are shown in Figure \ref{fig:epsilon-spacial-correlator}. The boosted frame results have significantly less volume dependence, and remain accurate across a wide range of $m$ and $r$. A larger boost improves the accuracy at higher $m$ provided that the cutoff $\Lambda$ is sufficiently large.

\begin{figure}[h!]
\begin{center}
\includegraphics[width=0.49\textwidth]{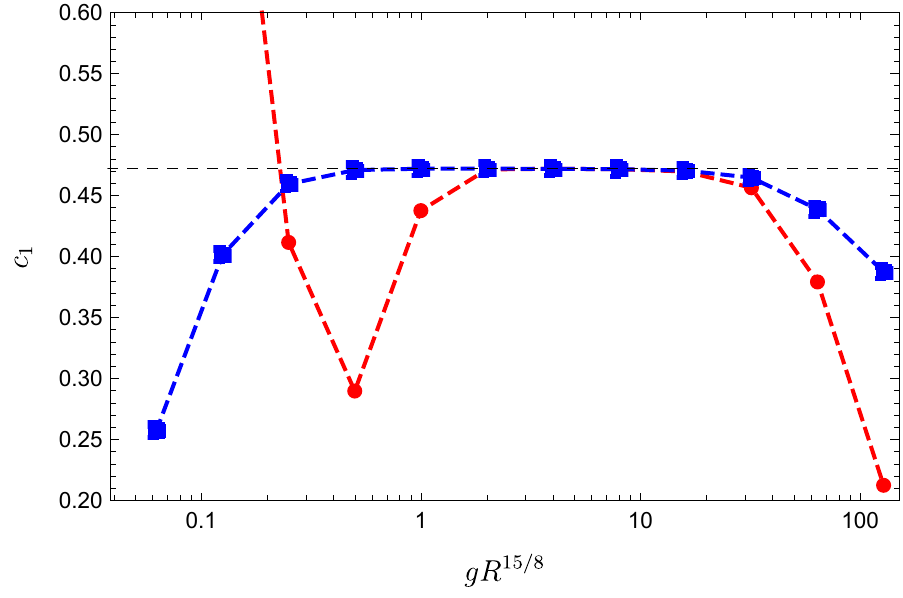}
\includegraphics[width=0.49\textwidth]{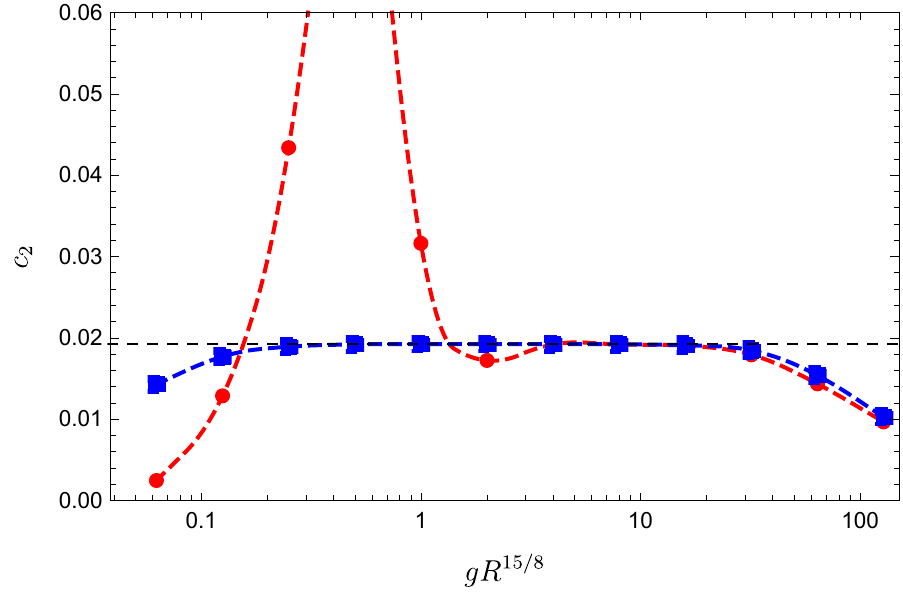}\\
\vspace{0.3cm}
\qquad\qquad\qquad\quad
\includegraphics[width=0.65\textwidth]{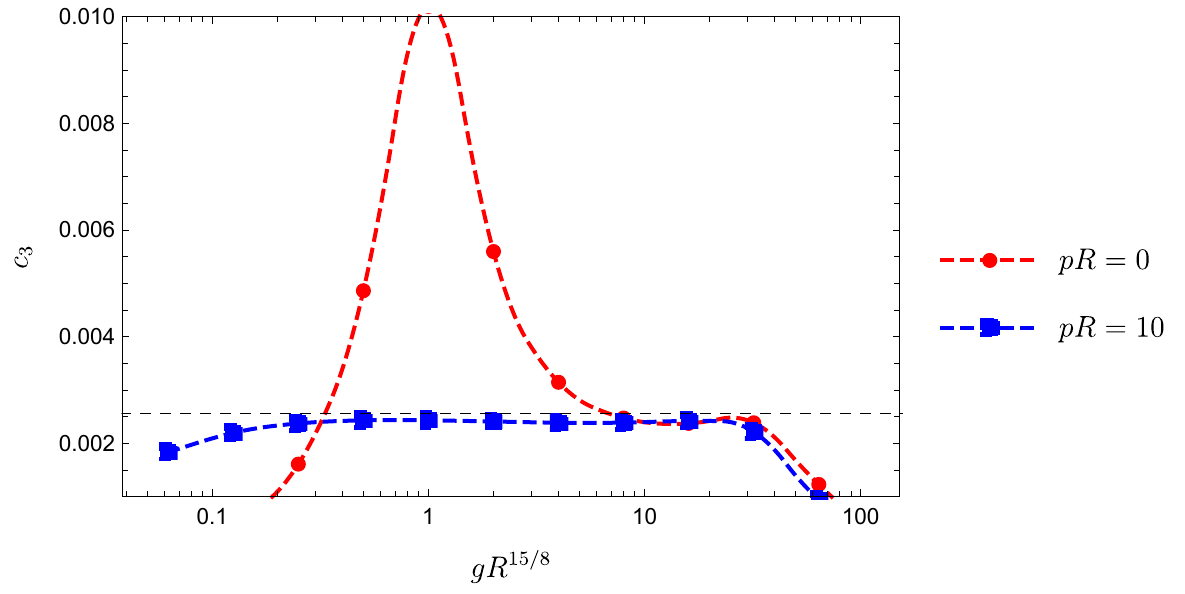}
\caption{Contributions to $c$-function from the first three $E_8$ states of the $\sigma$ deformation, comoputed using TCSA in the rest frame $pR=0$ and in a boosted frame with $pR=10$. The black dashed lines are the theoretical results in Table \ref{table:E8} from integrability  for comparison. }
\label{fig:sigmaDeformationDeltaCi}
\end{center}
\end{figure}

For the $\sigma$ deformation, the mass gap measured in the rest frame is given in Figure \ref{fig:sigmaDeformationMassGap}.   We then use the procedure outlined at the beginning of this section to compute the contributions to the $c$-function from the first three $E_8$ states in boosted frames. We compare the result of $c_i$ with $i=1, 2, 3$ in a boosted frame with those of the rest frame in Figure \ref{fig:sigmaDeformationDeltaCi}. One can see again that boosting improves the accuracy of the measurements significantly. In all three cases, the ranges of $g$ where the result is accurate are extended, especially in the small $g$ direction. The improvement is most striking for $c_3$. For the rest frame, $c_3$ only roughly agrees with the theoretical result for a small range of $g$ (specifically, $10\lesssim g R^{15/8} \lesssim 30$), however, for the boosted frame with $pR=10$, the agreement is good down to $g R^{15/8}\sim 0.2$. We also see that all three $c_i$s agree with the theoretical results in roughly the same range of $g$.

\subsection{Ising model deformed by the \texorpdfstring{$\sigma$}{sigma} and \texorpdfstring{$\varepsilon$}{epsilon} operators}\label{sec:isingFieldTheory}
So far, in this section, we have considered the two integrable limits $g=0,m\neq 0$ and $g\neq0,m=0$, and compared the spectra and $c$-functions from truncation with those of the exact solution. In this subsection, we consider the Ising field theory (IFT), where both deformations are turned on, that is, $g\neq 0$ and $m\neq0$.
For convenience, we parametrize the theory in terms of $\eta$ and $t$ defined as follows:
\begin{equation}
	\eta=\frac{m}{h^{8/15}},\quad t=\sqrt{m^2+h^{16/15}}\, ,\quad \textrm{with } g=2\pi h
\end{equation}
where $\eta$ determines the physics in the infinite volume limit, and $t$ can be considered as the volume of the system (in the units with $R=1$). Since the theory is not integrable, the general strategy of obtaining reliable infinite-volume measurement is by checking the consistency across a range of $\Lambda$ and $t$. For fixed $\eta$, $t$ controls the balance between convergence and finite-volume corrections. For small $t$, the result converges very quickly as $\Lambda$ gets big, but the $\Lambda\rightarrow \infty$ limit has a significant finite volume correction. On the other hand, for large $t$, the finite volume correction is exponentially suppressed, but the result is still varying at the values of $\Lambda$ that we are able to go up to, making it more difficult to extrapolate to $\Lambda\rightarrow \infty$.

\begin{figure}[htbp]
\centering
\includegraphics[width=0.5\linewidth]{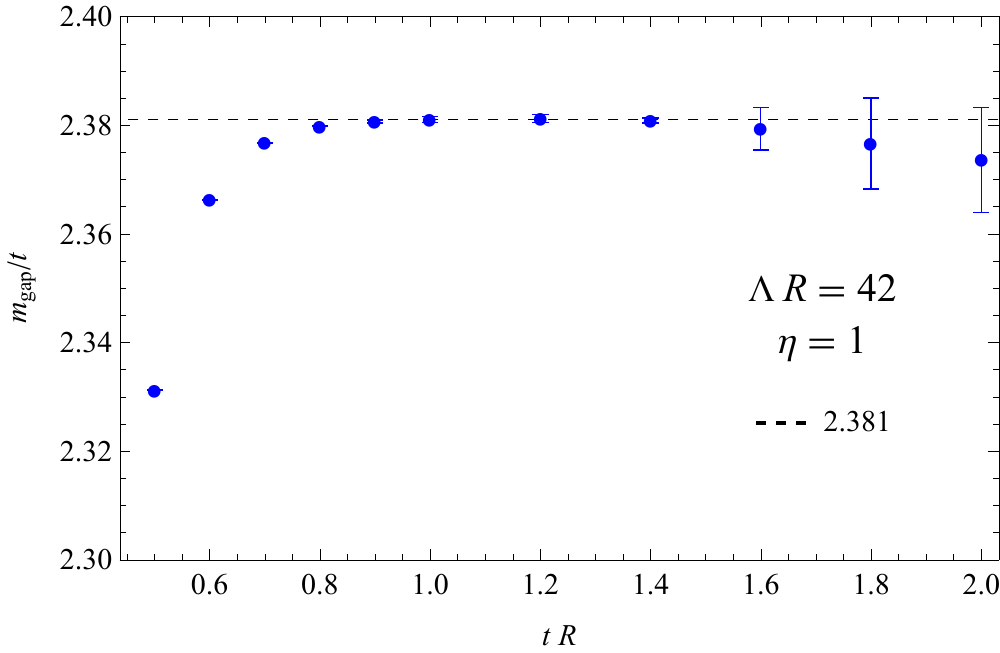}
\caption{
	\label{fig:gap-eta-1} 
Measurement of the relative mass gap $\frac{m_{\rm gap}}{t}$ at $\eta = 1$ as a function of $t$. Each point is obtained by fitting the data of certain $t$ as a quadratic polynomial of $\frac{1}{\Lambda R}$, and extrapolating to $\Lambda\rightarrow \infty$. The error bar is estimated by the difference between the extrapolated values from fitting models that are quadratic and cubic in $\frac{1}{\Lambda R}$. In the infinite volume $t\rightarrow \infty$ limit,  $\frac{m_{\rm gap}}{t}$ should approach a constant, but large $t$ data are subject to larger truncation errors. This constant (represented by black dashed line) is estimated from intermediate $t$ data where the values are consistent (with small errors) for a range of $t$.
}
\end{figure}
\begin{figure}[htbp]
\centering
\includegraphics[width=0.65\linewidth]{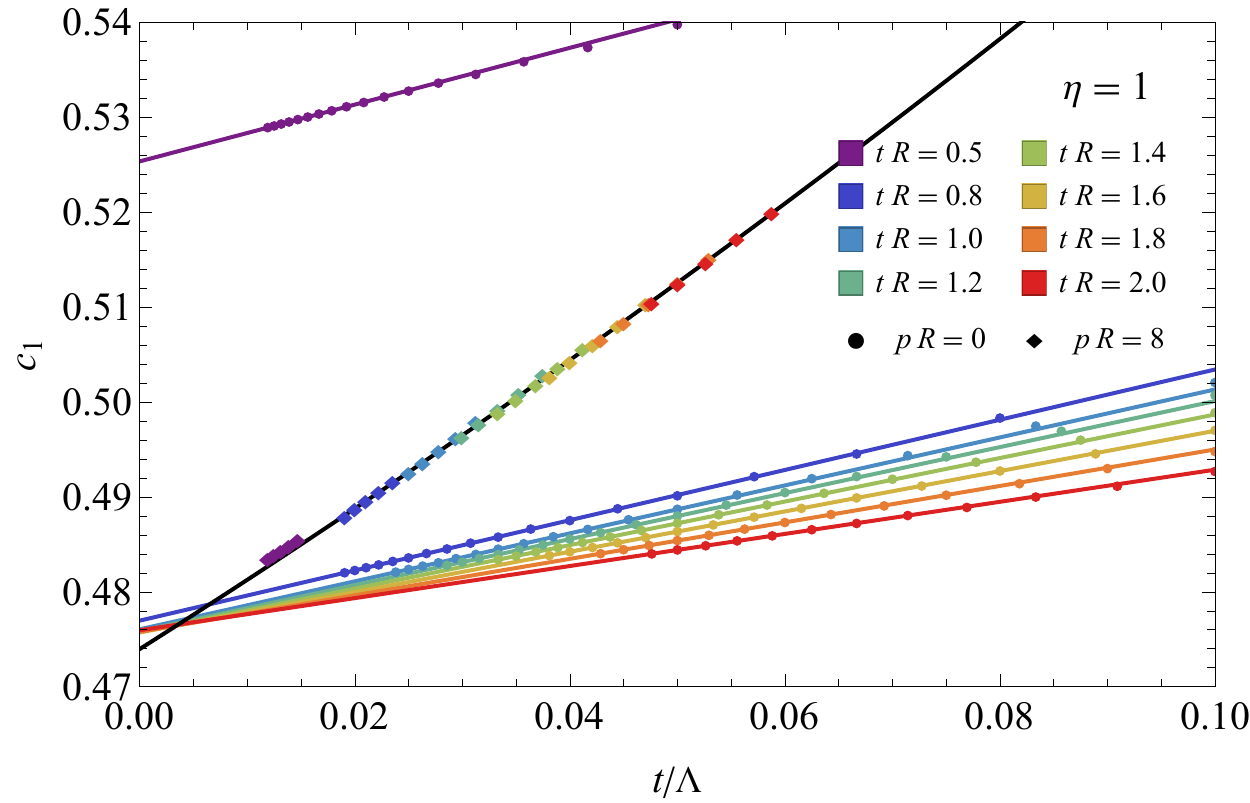}
\caption{
	\label{fig:c1-eta-1} 
	Measurements of $c_1$, the contribution to the $c$-function from the lowest bound state at $\eta=1$, as a function of the rescaled inverse cutoff $t/\Lambda$. The rest frame measurement is shown as circles, and the boosted result is shown as diamonds. The circles/diamonds represent the $c_1$ measurements made at different $t$ and $\Lambda$, and data at different $t$ are represented by different colors. The solid curves are obtained by fitting the data with a model of quadratic polynomial in $\frac{t}{\Lambda}$. For the rest frame result, while for $Rt > 0.5$, the curves extrapolate to approximately the same point, for finite $\Lambda$, the results of different $t$ (different colors) converge at different rates. In contrast, at $pR=8$, the results at different $t$ all join into a universal function of $\frac{t}{\Lambda}$.
}
\end{figure}
For a non-perturbative example, we take $\eta = 1$. In Figure \ref{fig:gap-eta-1}, we plot the mass gap obtained from TCSA in the rest frame. Similar to the previous sections, the measured mass gap is stable for a range of $t$ that is wide enough to convince us that we have measured the infinite-volume mass gap quite accurately.
Using this fit for the mass gap $\frac{m_{\rm gap}}{t} = 2.38$, we can compute the spectrum and the $c$-function using TCSA in  boosted frames. 
Like the $\eta = 0$ case, the lowest energy eigenstate is a bound state and has a stable finite contribution to the $c$-function in the infinite volume limit, denoted as $c_1$. 
In Figure \ref{fig:c1-eta-1}, we compare measurements of $c_1$ in the rest frame and the $pR=8$ boosted frame. We see interesting universal behavior in the boosted frame results, with the convergence of $c_1$ following the same function of $\frac{t}{\Lambda}$. We have seen this behavior in the mass gap fit of the $\eta\rightarrow \infty$ (that is, $g\rightarrow 0$) result in Figure \ref{fig:epsilon-rest}. We may interpret this universal behavior as the disappearance of the volume dependence at a large boost. In this case,  the convergence only depends on the dimensionless ratio of the UV cutoff $\Lambda$ and the deformation scale $t$. By contrast, at $p=0$, there is likely additional volume dependence that changes the rate of convergence.

\subsection{Phase transition across \texorpdfstring{$m=0$}{m=0}}

Spontaneous symmetry breaking and phase transitions are interesting phenomena to study with truncation. In this subsection, we explore the phase transition of the Ising field theory in boosted frames and see if there is any improvement. At $g=0$, the theory has a $\mathbb{Z}_2$ symmetry under which $\sigma \rightarrow -\sigma$ and $\varepsilon\rightarrow \varepsilon$. At $m < 0$, the symmetry is spontaneously broken, and two vacua are formed with $\<\sigma\>\neq 0$. From the broken phase, if we turn on a small $\sigma$ deformation, the symmetry will be explicitly broken and the false vacuum will be lifted with $E_{\rm false} - E_{\rm true}$ proportional to the volume. We thus have an interesting phase transition. At $\eta \gg 1$, the IFT is in the ``high-temperature'' phase containing weakly coupled massive fermions \cite{Zamolodchikov:2013ama}, while at $-\eta \gg 1$ (i.e. $\eta<0$ and $|\eta| \gg 1$) the IFT is in the low energy phase, where fermions are shown to confine into a tower of mesons \cite{Fonseca:2006au}.\footnote{In the dual bosonic picture, the low energy phase is a double-well potential for a scalar $\phi$, and the fermions are kink and anti-kink solitons that interpolate between the two vacua.  In this picture, the $\sigma$ deformation is a linear term $\sim \phi$ that splits the two vacua in energy and therefore causes kink and anti-kink solitons to confine.}
It is challenging to study the $-\eta \gg 1$ regime because it is difficult to resolve the various bound states at finite truncation. Moreover, lifting the false vacuum requires the volume factor $t$ to be large, which in turn requires much higher truncation $\Lambda$. 

\begin{figure}[htbp]
\includegraphics[width=0.45\linewidth]{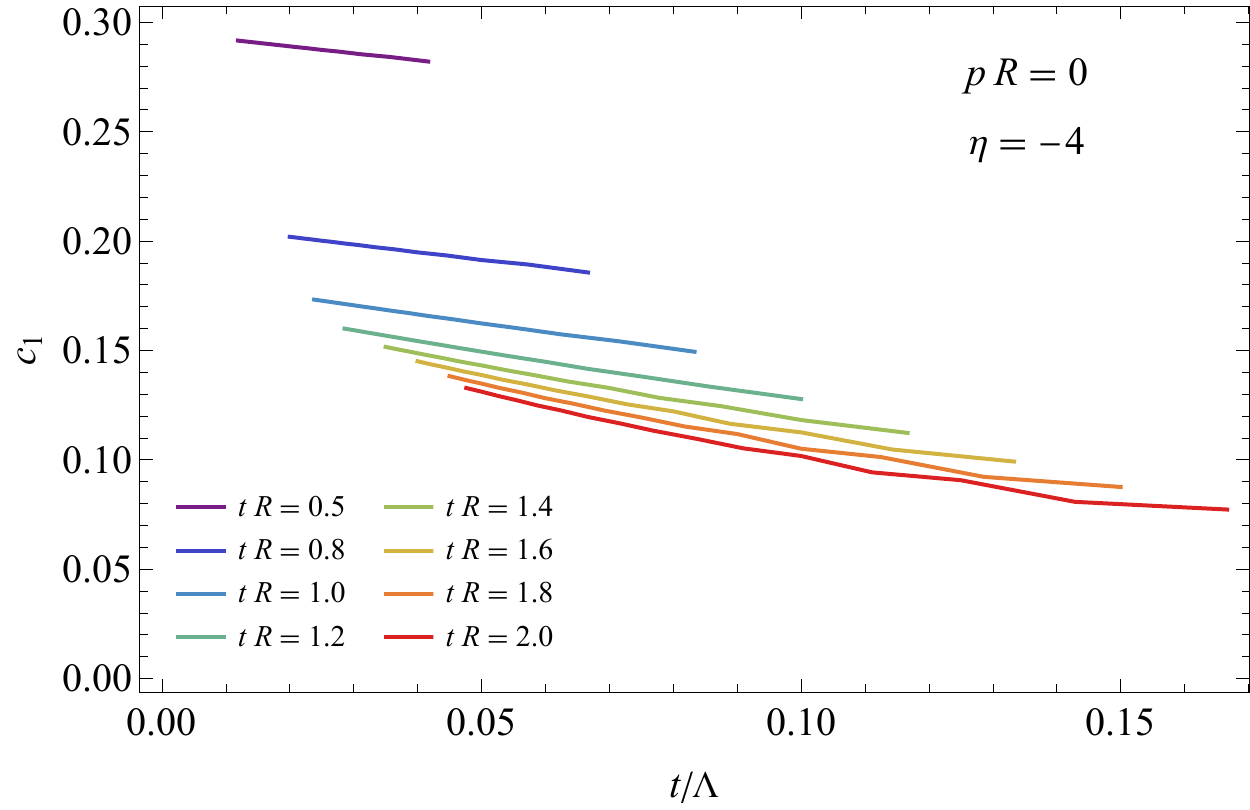}
\includegraphics[width=0.45\linewidth]{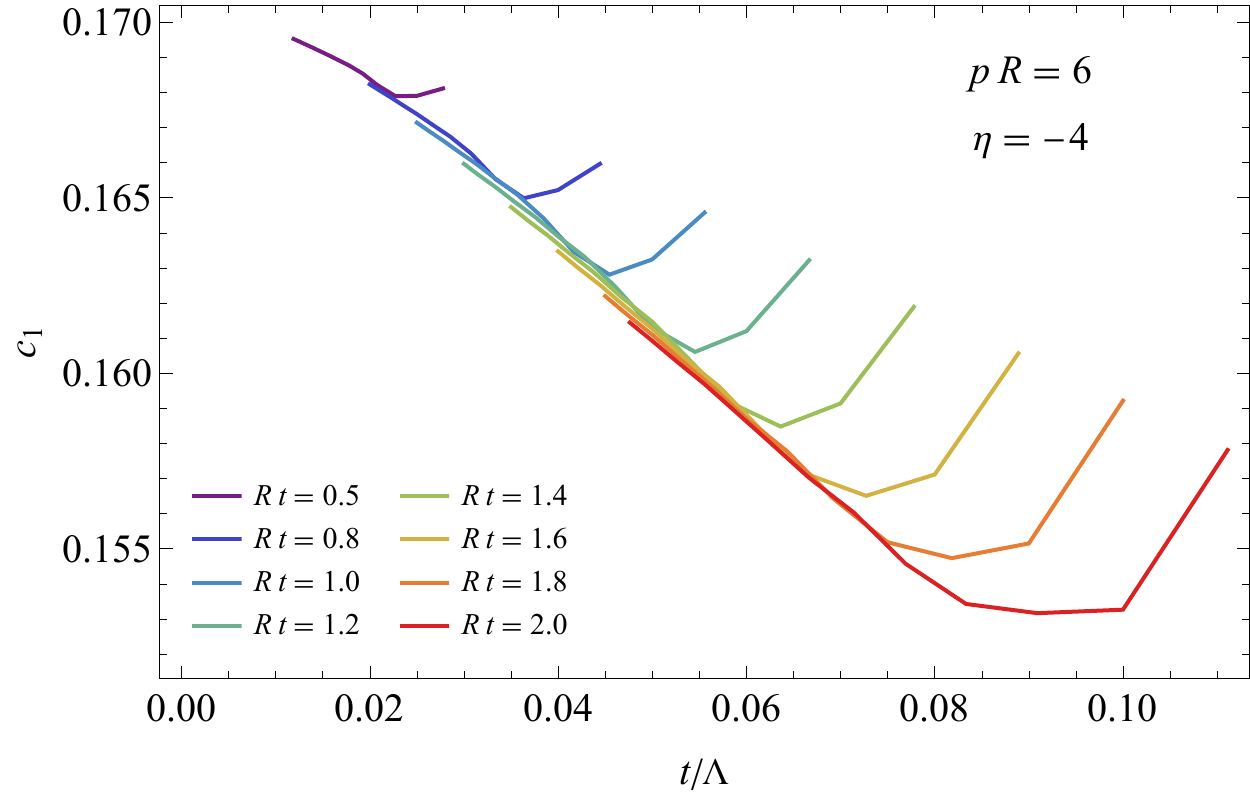}
\caption{
	\label{fig:c1-eta-minus4}
	The contribution to $c$-function from the lowest bound state, $c_1$. {\bf Left panel:} the rest frame measurement. The result is still converging, and has a strong dependence on volume. {\bf Right panel:} the boosted frame measurement with $pR=6$. 
}
\end{figure}
\begin{figure}[htbp]
\includegraphics[width=0.45\linewidth]{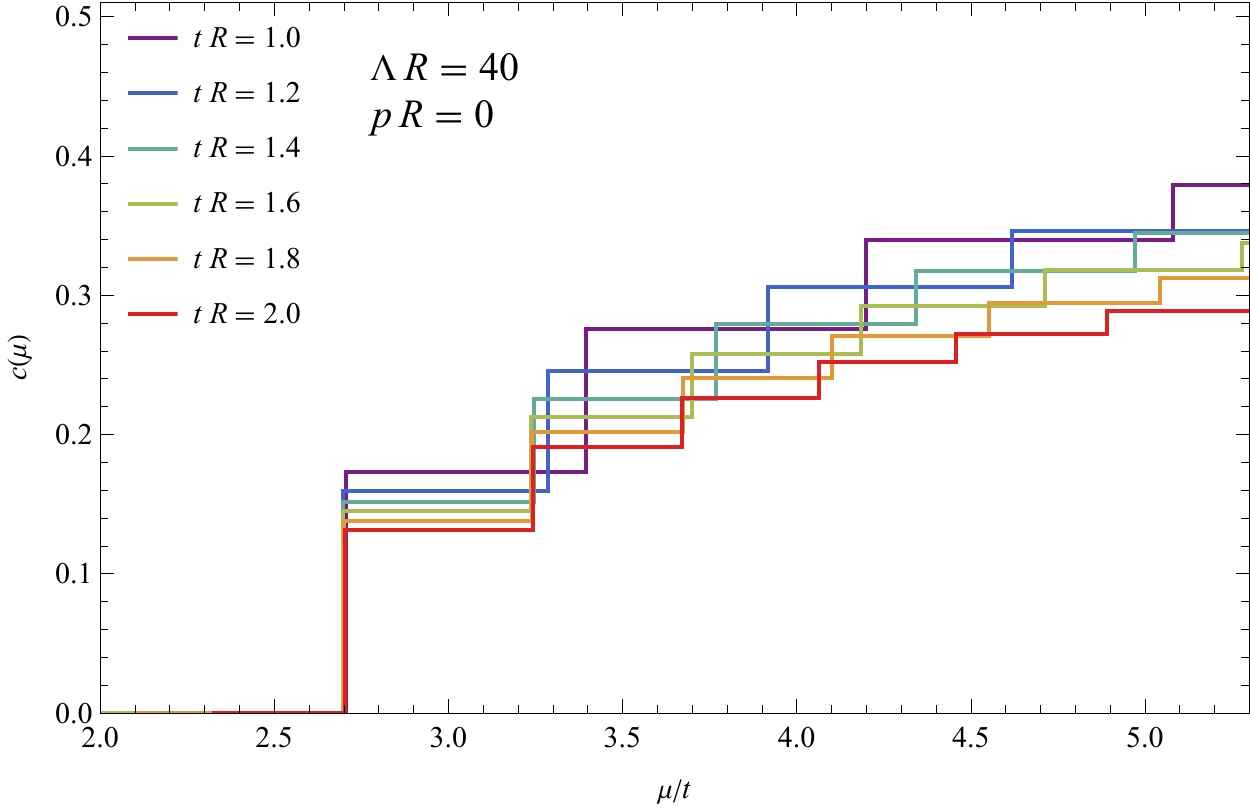}
\includegraphics[width=0.45\linewidth]{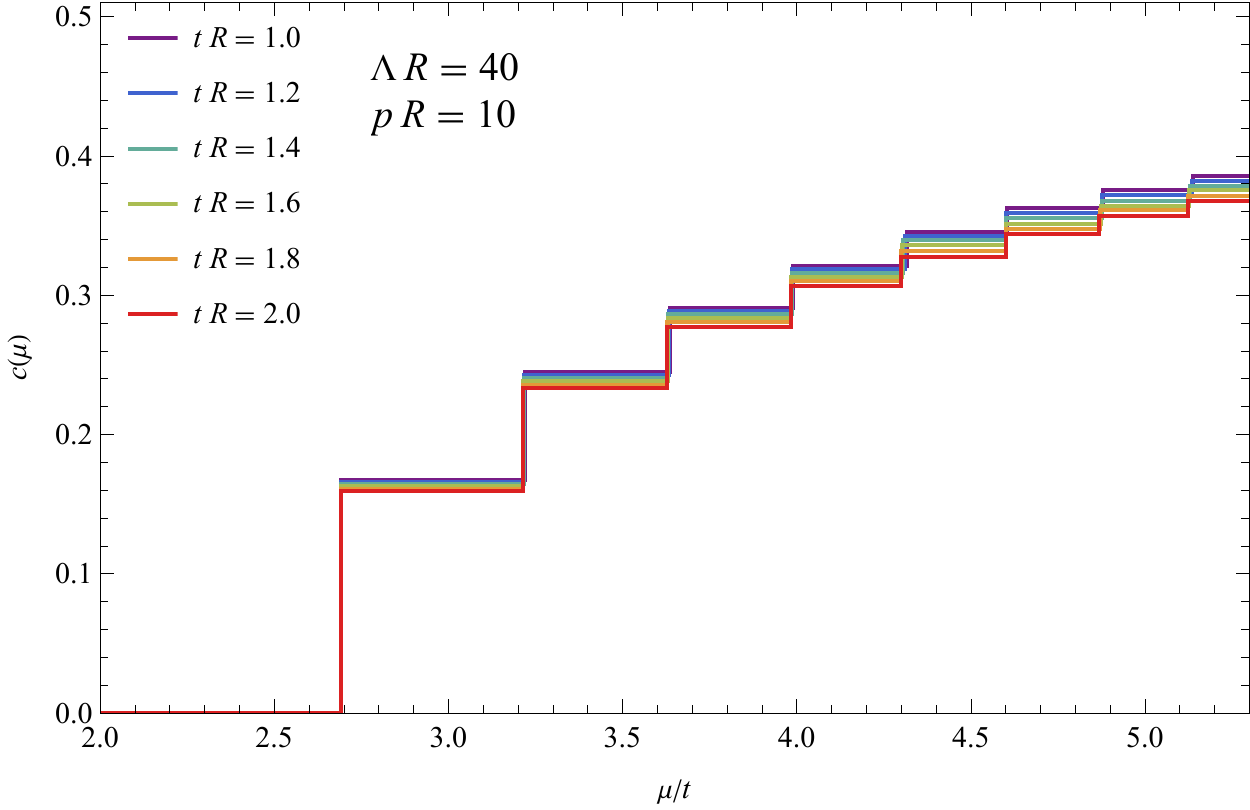}
\caption{
	\label{fig:c-func-eta-minus4}
	The $c$-function as a function of the invariant mass $\mu$ at $\eta = -4$. {\bf Left panel:} The rest frame measurement. {\bf Right panel:} The boosted frame measurement with $pR=10$.  In this case, the discrete steps are at the locations of the masses of stable particles that exist in the infinite volume limit, and we see little dependence on $m$ as expected. 
}
\end{figure}
To give an example of how difficult it is to study the $-\eta \gg 1$ regime in the rest frame, in the left panel of Figure \ref{fig:c1-eta-minus4}, we compute $c_1$ at\footnote{This value of $\eta$ is chosen such that the explicit $\mathbb{Z}_2$ breaking is small, but not too small to allow a finite-truncation measurement. For more negative $\eta$, the volume required for sufficiently lifting the false vacuum is significantly larger, which in turn requires significantly larger truncation $\Lambda$ to converge.} $\eta = -4$. Indeed, the measurement in the rest frame has a strong dependence on the $t$  which measures the volume, making it difficult to extract a stable result. In contrast, the $pR=6$ boosted frame result in the right panel of Figure \ref{fig:c1-eta-minus4} is consistent for $1.0 \leqslant tR\leqslant 2.0$. Therefore, in the $-\eta\gg 1$ regime, measuring the $c$-function in the rest frame is difficult, but it is still feasible with a sufficiently large boost. We also observe a big improvement in the shape of the $c$-function in  boosted frames, shown in Figure~\ref{fig:c-func-eta-minus4}. In the $-\eta\gg 1$ regime, we expect a series of stable bound states, whose masses $m_i$ and contributions to the $c$-function are expected to not change much for converged, volume-independent truncation. In Figure~\ref{fig:c-func-eta-minus4}, we see that the behavior of the boosted frame measurement agrees with this expectation, while the rest frame measurement still depends significantly on the volume.

\begin{figure}[htbp]
\includegraphics[width=0.45\linewidth]{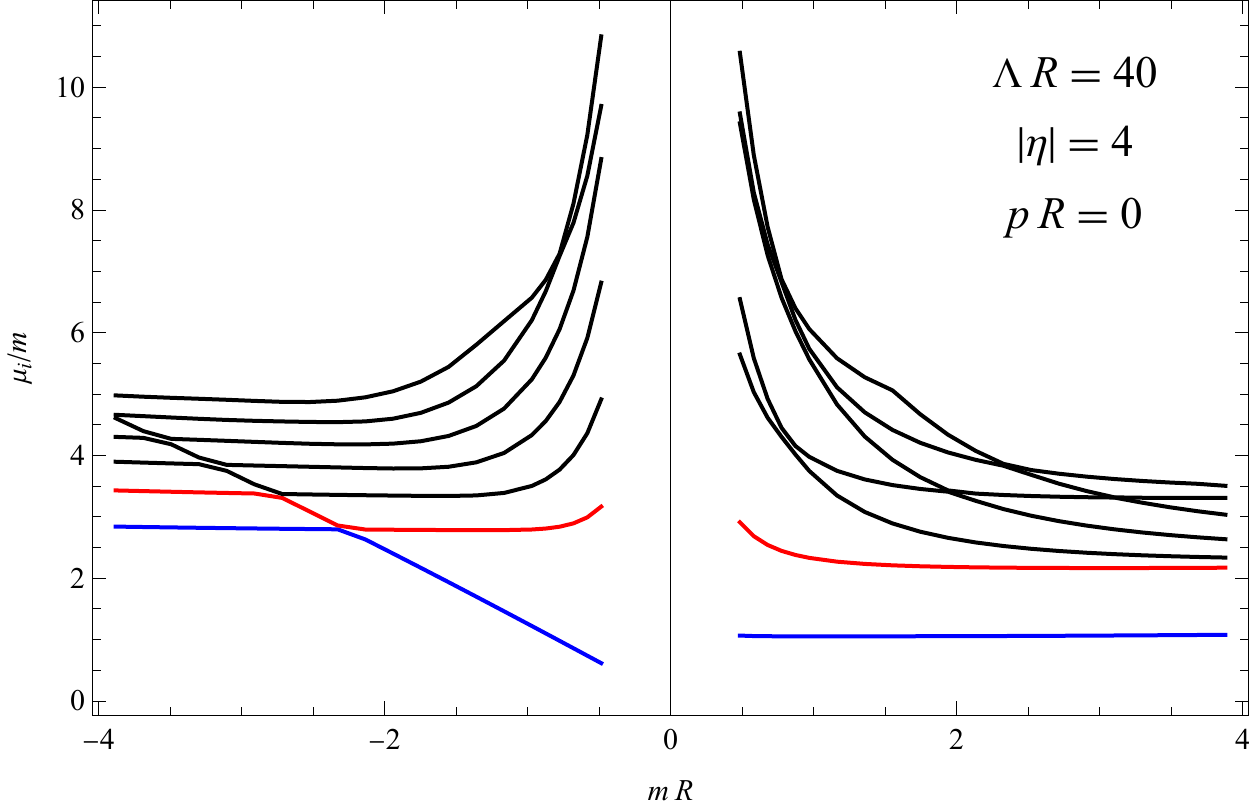}
\includegraphics[width=0.45\linewidth]{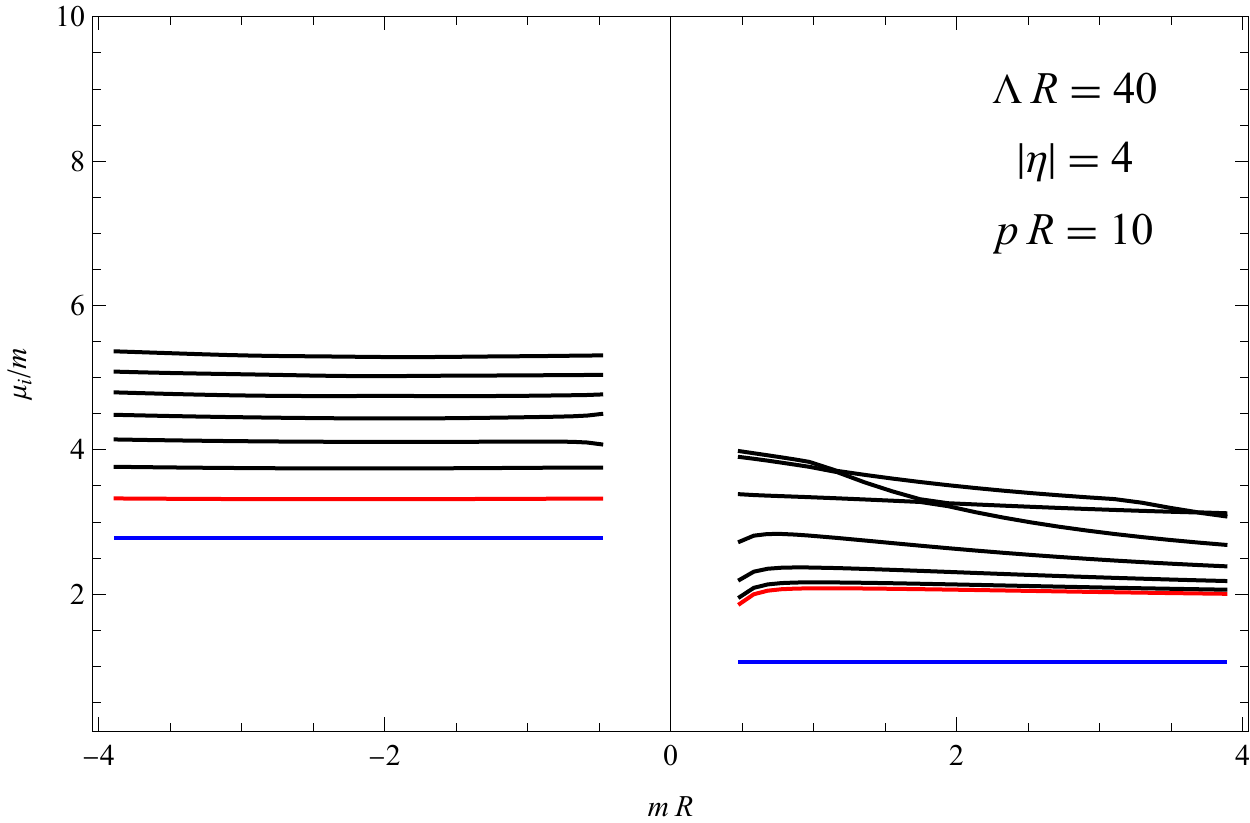}
\caption{
	\label{fig:spectrum-eta-pm4} The low-lying spectrum of $|\eta|=4$ IFT as a function of $m$. The lowest and second-lowest eigenvalues are colored blue and red, respectively. 
	{\bf Left panel:} The rest-frame spectrum. On the $m>0$ side, there is a stable single-particle state, followed by the multi-particle continuum states.
	On the $m<0$ side, the lowest eigenvalue scales with $m$, indicating it is the false vacuum state; the rest of the states shown in the plot are stable bound states.
	 {\bf Right panel:} The boosted $pR=8$ spectrum. On the $m>0$ side the physics is qualitatively the same as the $pR=0$ case, with less dependence on volume. On the $m<0$ side, the false vacuum state disappears, and the rest of the states have much less volume dependence.}
\end{figure}
To study the phase transition, we fix $|\eta|=4$ and scan $m\approx t$ from $mR\ll -1$ to $mR\gg 1$. The volume is controlled by $m$, so these two limits correspond to the infinite volume results in each phase. 
In Figure \ref{fig:spectrum-eta-pm4}, we show the low-lying spectrum as a function of $m$. We see the correct physics in both the rest frame measurement and the $pR=10$ boosted frame measurement. At $mR\gg 1$, the spectrum contains a single-particle state and a number of two-particle states, forming a quasi-continuum with an accumulation point at $2m_{\rm gap}$. When $m$ is reduced, the spectrum goes through a transition region, where the finite volume effect is significant. At $-mR\gg 1$, the spectrum completely reshuffles into bound states. However, there are significant differences between the $p=0$ and the $pR=10$ measurements. At $p=0$, we see the lowest eigenvalue is proportional to the volume $m$, and this behavior indicates that the corresponding state is the false vacuum, which is being lifted. At $pR=10$, the false vacuum state disappears from the beginning. Moreover, the volume dependence of the $pR=10$ results is much weaker than that in the rest frame.

\begin{figure}[htbp]
\centering
\includegraphics[width=0.45\linewidth]{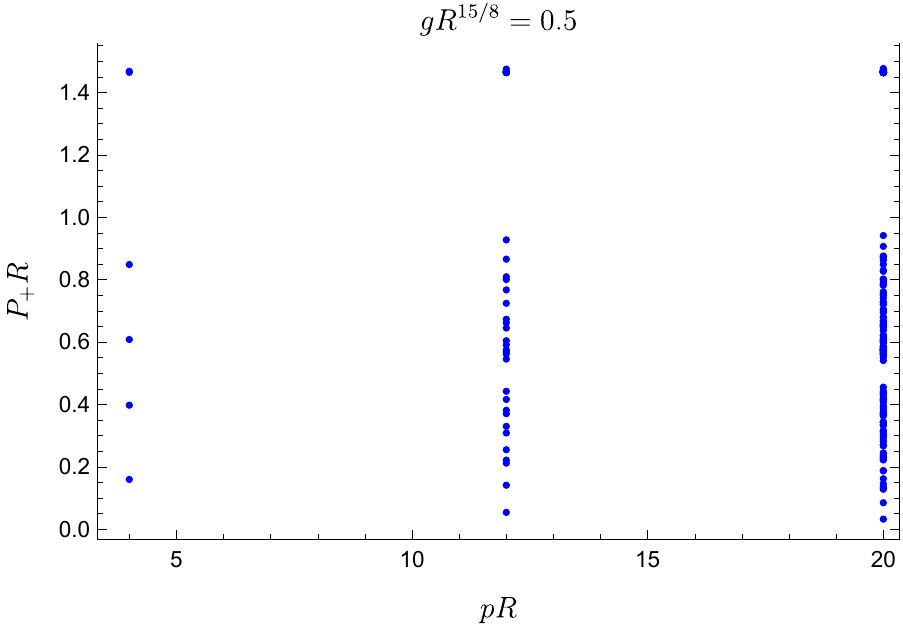}\qquad 
\includegraphics[width=0.45\linewidth]{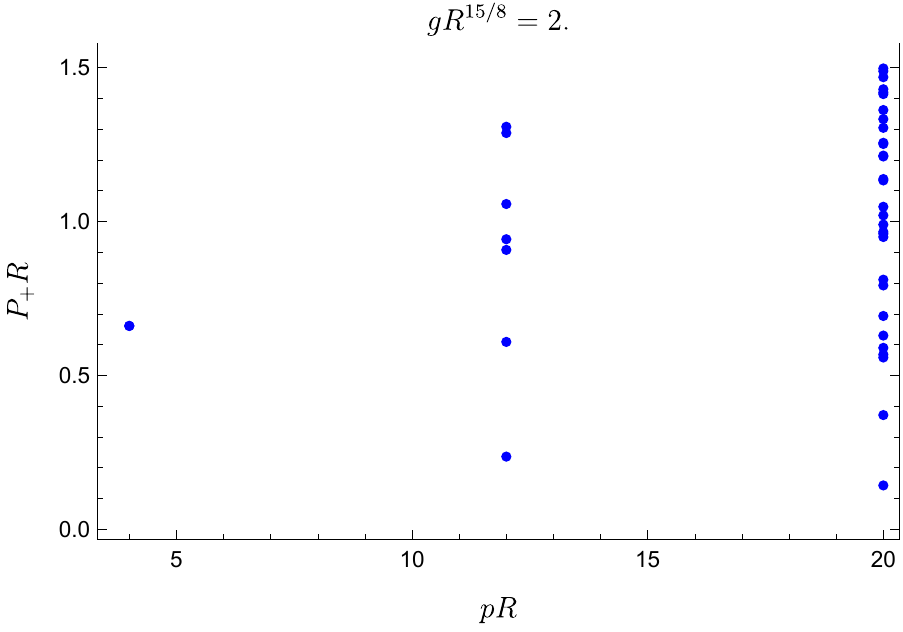}
\caption{The $P_+=H-p$ spectra for the $\sigma$ deformation for different values of $g$ and $p$. Each point in the plots represents one eigenvalue of the corresponding $P_+$. We used $\Lambda=40$ when making these plots.
}
\label{fig:PplusSpectrumComparison}
\end{figure}

\begin{figure}[htbp]
\centering
\includegraphics[width=0.45\linewidth]{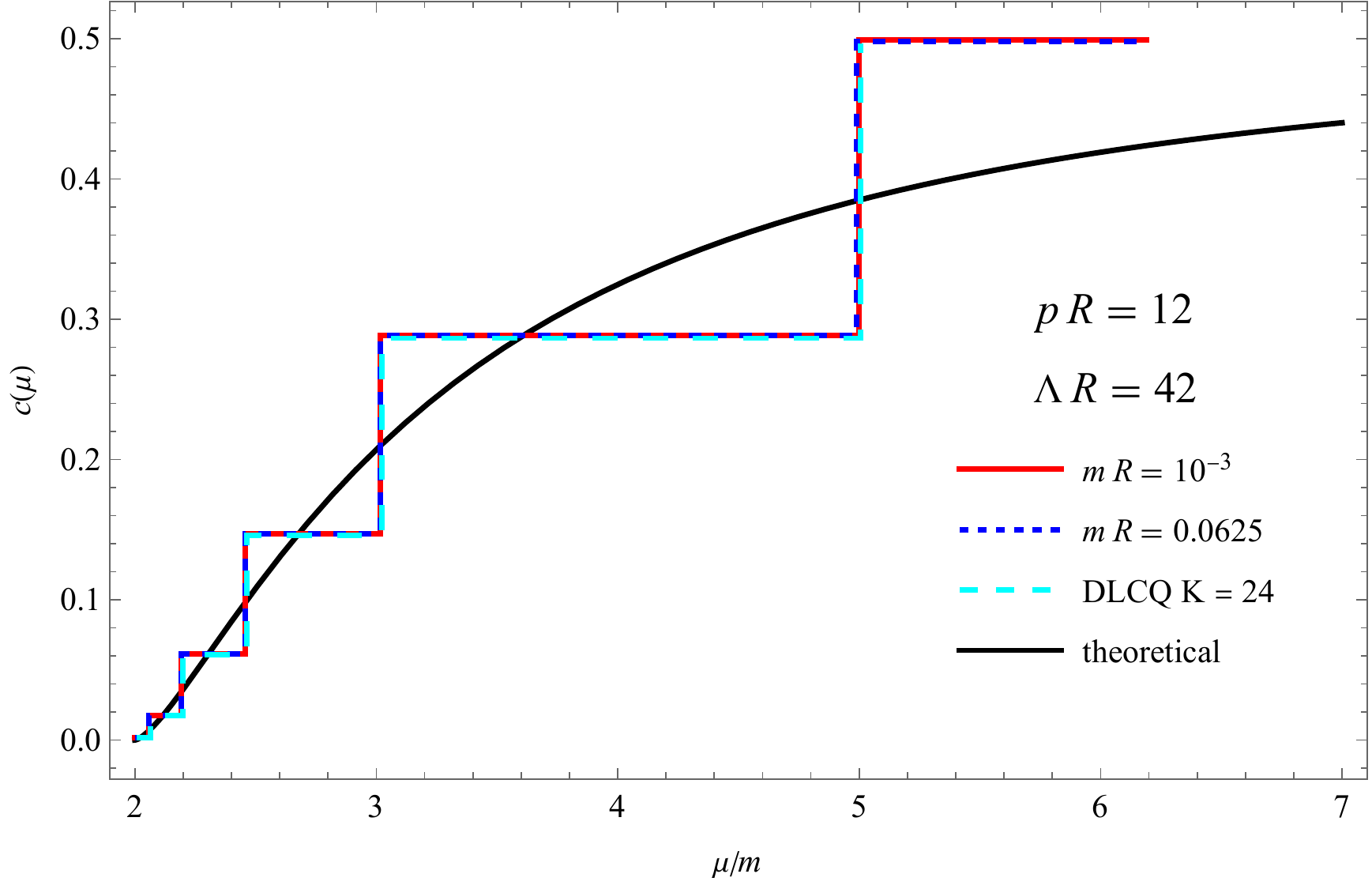}
\caption{
The $c$-function for the $\varepsilon$ deformation at small values of $m$. The truncation results (red and blue) measured in a boosted frame with $pR=12$ agree with the Discrete Lightcone Quantization (DLCQ) result (green), which is the answer expected at infinite boost.
}
\label{fig:compareWithDLCQ}
\end{figure}

\section{Discussion}
\label{sec:Discussion}

In this work we have seen that boosting can improve the results and lessen the dependence on volume, thereby coming closer to approximating the infinite volume answer.
At the same time, lightcone (LC) quantization is an approach that keeps the vacuum trivial and therefore has no dependence on the volume.  This suggests that even at finite moderate boosts, there is a set of low-energy effective degrees of freedom that are increasingly behaving as if they were quantized on the lightcone.  Examining the spectrum of the LC Hamiltonian $P_+\equiv H-p$ for the $\sigma$ deformation, which we show in Figure \ref{fig:PplusSpectrumComparison}, we see states accumulating near $P_+=0$.  Indeed, this is the behavior on the lightcone ($p\rightarrow \infty$), where $P_+^{(n)} = m_n^2/(2p)$.  The naive expectation is that the $P_+$ low-energy degrees of freedom are simply the chiral modes, that is the multi-particle states which have $P_+=0$ before adding the $\sigma$ deformation.  In other words, the effective theory is generated by integrating out all states which had $P_+ > 0$ before adding the relevant deformation.  The induced LC effective Hamiltonian $P_+^{\rm eff}$, then acts solely on the chiral states and should have approximately the same spectrum as the low energy spectrum of the full $P_+$ that acts on the whole Hilbert space.  This interpretation is consistent with the phenomena we observe upon boosting.  In particular, we find that the results become worse if $p  \gtrsim \Lambda$, which follows from the fact that the induced LC Hamiltonian is obtained from integrating out modes with $P_+ \sim p$, which thus must be included in the truncation to obtain the correct induced terms.  In addition, we can understand the difference in the range of $m$ vs. $g$ needed to obtain good results.  The $\varepsilon$ deformation induces an effective Hamiltonian at ${\cal O}(m^2)$, and therefore even if $mR \rightarrow 0$, so that perturbation theory in $m$ is reliable, one gets the correct LC behavior (see Figure \ref{fig:compareWithDLCQ}).  On the other hand, for the $\sigma$ deformation, the LC Hamiltonian must necessarily be non-perturbative in the coupling by dimensional analysis (since the mass dimension of $g$ is $[g]=15/8$), which implies that when $g$ is small enough for perturbation theory to be valid, the LC effective Hamiltonian will not be properly induced (note the small coupling behavior in Figure \ref{fig:sigmaDeformationDeltaCi}).

  Given the above indirect evidence for an effective LC Hamiltonian, a natural question is whether there is a direct formula for computing it from boosted TCSA data?
We will provide an explicit non-perturbative prescription for computing $P_+^{\rm eff}$ in a future paper \cite{effH}.  Note, that such a $P_+^{\rm eff}$, which in general depends non-perturbatively on the relevant couplings, does not mix the Fock vacuum with any chiral states.  Formally, it thus has no volume dependence, and is invariant under boosts.  However, in practice, if it is non-perturbative in the couplings (such as for the $\sigma$ deformation), it will only be possible to numerically compute it for some range of volumes at finite truncation.  Finally, let us note that since the spectrum of $P_+^{\rm eff}$ approximately matches the spectrum of the full $P_+$, it will also be able to accommodate a phase transition.  In fact, it will be discontinuous across the phase transition, in a manner similar to the right panel of Figure \ref{fig:spectrum-eta-pm4}.

\newpage
\begin{center}
\subsection*{Acknowledgments}
\end{center}
We thank Matthew Walters for helpful conversations. HC, ALF, and EK were supported in part by the Simons Collaboration Grant on the Non-Perturbative Bootstrap, and ALF in part by a Sloan Foundation fellowship. YX was supported by a Yale Mossman Prize Fellowship in Physics.

\

\appendix

\section{Technical Details of TCSA for the Ising Field Theory}\label{app:TCSA}
In this paper, we studied the thermal and magnetic deformations of the 2d Ising CFT (free fermion) using TCSA. In this appendix, we provide some technical details about how to compute the $\sigma$ and $\varepsilon$ matrices. 

 To be specific, the Hamiltonian that we are studying on a cylinder of radius $R$ is given by
\begin{equation}\label{eq:AppendixHamiltonian}
	H=H_{0}+\frac{1}{2\pi}\int_{0}^{2 \pi R} d x \,\big(  m\varepsilon(x) + g\sigma(x) \big) \, .
\end{equation}
where $H_{0}$ is the free fermion Hamiltonian
\begin{equation}
	H_{0}=\frac{1}{2 \pi} \int_0^{2\pi R}dx(\psi \bar{\partial} \psi+\bar{\psi} \partial \bar{\psi})
\end{equation}
The Hiblert space of $H_0$ is a direct sum of the Neveu-Schwarz (NS) and Ramond (R) sectors, 
\begin{equation}
\mathcal{H}^{\text {free }}=\mathcal{H}_{\mathrm{NS}} \oplus \mathcal{H}_{R}.	
\end{equation}
The NS sector has a unique vacuum $|0\rangle$ and states are built up by acting on the vacuum with fermionic operators $a_{n+1/2}$ and $\bar a_{n+1/2}$ ($n\in \mathbb{Z}$). The R sector has degenerate vacua $|\sigma\rangle$ and $|\mu\rangle$, and the exited states are obtained by acting with even number of the fermionic operators $a_n$ and $\bar a_n$ ($n\in\mathbb{Z}$) on the corresponding vacua (odd number of fermionic operators switches the vacua). These fermionic operators are the coefficients in the mode expansions of the $\psi$ and $\bar\psi$ (for more details, see equaitons (2.3) and (2.4) of  \cite{Yurov:1991my}), and they obey the usual anti-commutation relation $\{a_i, a_j\}=\delta_{i+j}$, and similarly for $\bar{a}_i$.

For the interest of this paper, we consider the NS vacuum $|0\rangle$ and the R sector $|\sigma\rangle$ vacuum, and states generated with even number of fermionic operators acting on these vacua. Specifically, we have \footnote{Note that in the states in the R sector, if $a_0$ or $\bar{a}_0$ appears, then it should be understood that each $a_0$ or $\bar{a}_0$ comes with a factor of $\sqrt{2}$, such that the states in (\ref{eq:states}) all have norm equal to 1. This is implicit in equation (\ref{eq:sigmaME}).}
\begin{align}\label{eq:states}
&\bar{a}_{\bar{n}_{1}}...\bar{a}_{\bar{n}_{\bar{N}}}a_{n_{1}}...a_{n_{N}}|0\rangle,\quad \textrm{with } n_{i},\bar{n}_{i}\in\mathbb{Z}+1/2,\\
&\bar{a}_{\bar{n}_{1}}...\bar{a}_{\bar{n}_{\bar{N}}}a_{n_{1}}...a_{n_{N}}|\mathbb{\sigma}\rangle,\quad \textrm{with } n_{i},\bar{n}_{i}\in\mathbb{Z}\nonumber
\end{align}
where $n_{i},\bar{n}_{i}\le0 $ and $N+\bar{N}\in2\mathbb{Z}$. From a 2d CFT point of view, these would correspond to the Virasoro primary states $|0\rangle,  |\varepsilon\rangle, |\sigma\rangle$, and their Virasoro descendants. These states are the eigenstates of $H_0$, and their eigenvalues (which are their energies) are given by 
\begin{equation}
E=E_{0}+\sum_{i=1}^{N}\frac{n_{i}}{R}+\sum_{i=1}^{\overline{N}}\frac{\overline{n}_{i}}{R},
\end{equation}
where $E_0=0$ in the NS sector and $E_0R=1/8$ in the Ramond sector. Therefore, the $H_0$ matrix in this basis is simply a diagonal matrix with matrix elements given by the above formula. The momenta of these states are 
\begin{equation}
	p=\sum_{i=1}^{N}\frac{n_{i}}{R}-\sum_{i=1}^{\overline{N}}\frac{\overline{n}_{i}}{R}.
\end{equation}

For TCSA in the rest frame that is studied in section \ref{sec:RestFrameTCSA}, we simply include only states with $p=0$ and  $E\le\Lambda$ where $\Lambda$ is some energy cutoff of our choice. For TCSA in a boosted frame with momentum $p$ studied in section \ref{sec:BoostTCSA}, the basis consists of states with momentum $p$ and satisfying $E^2-p^2\le \Lambda^2$.

In terms of the fermionic fields, $\varepsilon$ is given by 
\begin{equation}
	\varepsilon=i\bar\psi\psi
\end{equation}
and its matrix elements in the states (\ref{eq:states}) can be simply computed by contractions of the fermionic generators. Note that the OPE coefficient of $\sigma\sigma\varepsilon$ is $1/2$, and this fixes $R^{1/8}\langle\sigma| \bar{a}_0 a_0\sigma|0\rangle$ to be equal to $-i/2$ (in \cite{Yurov:1991my}, they only considered the $\sigma$ deformation, and they used the convention that $R^{1/8}\langle\sigma| \bar{a}_0 a_0\sigma|0\rangle=1/2$). The computation of the $\varepsilon$ matrix is straight forward in the NS sector, while in the R section, there are some subtitles involving $a_0$ and $\bar{a}_0$ that need to be taken care of.

The matrix elements for the $\sigma$ operator are a little bit more complicated. One way to compute it is by taking the massless limit of the formula of the $\sigma$ matrix elements in the massive fermion basis \cite{Fonseca:2001dc}, and the result is given by (for the holomorphic part)\footnote{If $K+N$ is odd, then the full matrix element (including the anti-holomorphic part) must be proportional $\langle\sigma| \bar{a}_0 a_0\sigma|0\rangle$. The appearance of the factor $\sqrt{-i}^{(K+N)\text{ mod }2}$ is because in our convention $\langle\sigma| \bar{a}_0 a_0\sigma|0\rangle=-i/2$ and we distribute the $-i$ evenly in the holomorphic and anti-holomorphic parts. }
\begin{footnotesize}
\begin{equation}
\label{eq:sigmaME}
\left\langle \sigma\left|a_{n_{1}}\ldots a_{n_{K}}\sigma(0,0)a_{n_{K+1}}\ldots a_{n_{K+N}}\right|0\right\rangle =\frac{i^{K}\sqrt{-i}^{(K+N)\text{ mod }2}}{\sqrt{2\pi}^{N+K}R^{1/8}}\frac{\prod_{i=1}^{K}\kappa\left(n_{i}\right)/n_{i}}{\prod_{i=1}^{N}\kappa\left(\left|n_{K+i}\right|\right)}\left(\prod_{i=2}^{K+N}\prod_{j=1}^{i-1}\frac{n_{j}-n_{i}}{n_{i}+n_{j}}\right)
\end{equation}
\end{footnotesize}
where $n_i\ge0$ and $n_i\in \mathbb{Z}$ for $1\le i\le K$, while $n_i<0$ and $n_i\in \mathbb{Z}+1/2$ for $K< i\le K+N$. The function $\kappa(n)$ is given by\footnote{If $n_i=0$ in equation (\ref{eq:sigmaME}), then the factor $\kappa(n_i)/n_i$ should be understood as taking the $n_i\rightarrow 0$ limit, and we have $\lim_{n_i\rightarrow 0}\kappa(n_i)/n_i=\sqrt{\pi}$. } 
$\kappa\left(n\right)=\Gamma\left(n+1/2\right)/\Gamma(n)$. The anti-holomorphic part of the $\sigma$ matrix is given similarly, with one extra minus sign for each $\bar {a}_n$ for $n\in\mathbb{Z}$ (that is, for each anti-holomorphic mode of the Ramond sector). And when computing the full matrix elements, there may be an extra minus sign coming from anti-commuting the $a_{n}$s and $\bar{a}_n$s such that the full matrix elements factorize into a product of holomorphic and anti-holomorphic parts. A Mathematica notebook containing some simple code for computing the TCSA Hamiltonian of the IFT in the rest frame and boosted frames using the formulas given in this appendix is attached along with this paper.

However, a faster way of computing the $\sigma$ matrix elements in the states \eqref{eq:states} is to transform the fermionic generators $a_n$ and $\bar{a}_n$ to the periodicity-switching basis that is developed in \cite{Yurov:1991my}, in which the $\sigma$ matrix is simple to compute.  We attached a Mathematica package making use of the periodicity-switching basis for computing the $\sigma$ matrix, and a demo notebook about how to use it. This package is significantly fater than the simple code using formulas (i.e., \eqref{eq:sigmaME}) in this appendix, and it is used to compute all the results in this paper.

\bibliographystyle{JHEP}
\bibliography{refs}

\end{document}